\documentclass[a4paper,fleqn,usenatbib]{mnras}
\usepackage[T1]{fontenc}
\usepackage{ae,aecompl}
\usepackage{amsmath}
\usepackage{amsfonts}
\usepackage{amssymb}
\usepackage{graphicx}
\def\mnras{MNRAS}
\def\aap{A$\&$A}
\def\nat{Nature}
\def\apj{ApJ}
\def\apjl{ApJ Letters}
\def\apjs{ApJS}
\def\pasp{PASP}

\def\icarus{Icarus}

\newcommand{\mmsn}{{\rm MMSN}}
\newcommand{\me}{\, {\rm M}_{\oplus}}
\newcommand{\au}{\, {\rm au}}
\title[Giant planet formation in structured discs]{Giant planet formation in radially structured protoplanetary discs}
\author[G. A. L. Coleman and R. P. Nelson]{Gavin A. L. Coleman\thanks{Email: g.coleman@qmul.ac.uk} \& Richard P. Nelson \\
Astronomy Unit, Queen Mary University of London, Mile End Road, London, E1 4NS, U.K.}
\date{}
\begin{document}
\maketitle
\begin{abstract}
Our recent N-body simulations of planetary system formation, incorporating models for the main physical processes thought to be important during the building of planets (i.e. gas disc evolution, migration, planetesimal/boulder accretion, gas accretion onto cores, etc.), have been successful in reproducing some of the broad features of the observed exoplanet population (e.g. compact systems of low mass planets, hot Jupiters), but fail completely to form any surviving cold Jupiters. The primary reason for this failure is rapid inward migration of growing protoplanets during the gas accretion phase, resulting in the delivery of these bodies onto orbits close to the star. Here, we present the results of simulations that examine the formation of gas giant planets in protoplanetary discs that are radially structured due to spatial and temporal variations in the effective viscous stresses, and show that such a model results in the formation of a population of cold gas giants. Furthermore, when combined with models for disc photoevaporation and a central magnetospheric cavity, the simulations reproduce the well-known hot-Jupiter/cold-Jupiter dichotomy in the observed period distribution of giant exoplanets, with a period valley between 10--100 days.
\end{abstract}

\section{Introduction}
Ever since the discovery of the first extrasolar giant planet around a main sequence star \citep{MayorQueloz}, questions have been asked as to the formation and evolution of giant exoplanets. To date over 1640 confirmed extrasolar planets have been discovered, displaying a broad range of orbital and physical properties, and approximately 600 of these are believed to be gas giants \citep{exoplanets_org}. Explaining the origins of the broad diversity of exoplanets remains a formidable challenge to planet formation theory,  and even the more restricted challenge of explaining the orbital period distribution of giant exoplanets has not yet been addressed satisfactorily.

Observational biases, in particular the fact that ground based transit surveys are only sensitive to detecting giant planets with orbital periods $\lesssim 10$ days, and that radial velocity searches have surveyed stars that are more metal-rich than the average, give the impression that hot Jupiters are common. Recent studies that have examined data from the \emph{Kepler} spacecraft and follow-up radial velocity measurements \citep{Fressin13,Santerne16} find that hot Jupiters are expected to orbit only 1\% of main sequence stars, while cold Jupiters have a higher occurrence rate of 17\% \citep{Cassan12}.
The occurrence rate between the two populations does not increase linearly, however, as a `period valley' exists between 10--85 days where there is a dearth of giant planet detections when accounting for observational biases. This period valley was first observed in radial velocity surveys \citep{Udry03,Cumming2008}, and its existence has been supported by the aforementioned recent analysis of combined \emph{Kepler} and radial velocity observations \citep{Santerne16}. Individual theories have been put forward to explain this period valley \citep{HasegawaPudritz11,AlexanderPascucci12,ErcolanoRosotti15}, but none have been incorporated into ab initio models of planet formation to examine whether or not it is possible to explain this, and other features in the giant planet distribution, from first principles.

Competing theories of giant planet formation, including the core-accretion and pebble-accretion models \citep[e.g.][]{IdaLin2004,Alibert2006,Mords09,Bitsch15}, and the tidal-downsizing model \citep[e.g.][]{Nayakshin15}, have been used to make predictions about the giant planet population for comparison with observations, and to examine the formation of giant planets in our Solar System \citep{Levison15}. The fact that many multiplanet systems have been discovered, containing various combinations of super-Earths, Neptunes and Jovian mass bodies \citep{Muirhead2012, Neveu-VanMalle16,Becker15}, often in compact systems \citep[e.g.][]{Lissauer2011, Becker15}, suggests that gravitational interactions, and perhaps competitive accretion, are essential components of the planet formation process. Furthermore, the fact that many giant planets appear to be on eccentric orbits suggests that dynamical instabilities involving initially compact giant planet systems, either during or after formation, are common and important for shaping planetary system architectures \citep[e.g.][]{RasioFord1996}. For this reason, we have adopted the approach of using N-body simulations, combined with prescriptions for the major physical processes thought  to occur during planet formation, to simulate the building and early evolution of planetary systems. 

In recent work \citep{ColemanNelson14,ColemanNelson16}, we have examined the formation of planets in irradiated, viscous disc models that have adopted the standard $\alpha$ prescription \citep{Shak, Lynden-BellPringle1974}.
Except for a small region close to the star where the temperature exceeds 1000~K, we have assumed that $\alpha$ is constant, leading to smooth temperature and surface density profiles in the discs. The models have been successful in forming systems containing hot Jupiters, multiple super-Earths and Neptunes in compact configurations, and numerous terrestrial planets with a variety of compositions, but the models fail completely to form any surviving cold Jupiters. The main reason for this is that giant planet cores undergo rapid inward migration as they accrete gas, because the corotation torques that are needed to counteract the Lindblad torques become saturated \citep[e.g.][]{pdk11}. These planets then end up as hot planets orbiting close to the star. \cite{ColemanNelson14} undertook a detailed examination of the conditions required for giant planet formation and survival, and showed that a Jovian mass planet that settles into a final orbit at $5\au$ must have initiated runaway gas accretion and type II migration when at an orbital radius $\sim 15\au$, and this should have occurred late in the disc lifetime so that the gas disc disperses before the planet type II migrates all the way to the central star (or into the magnetospheric cavity if one is present). In this paper, we address the question of whether or not radial structuring of a protoplanetary disc, because of spatial and temporal variations in the viscous stress, can prevent accreting giant planet cores from migrating inwards rapidly because of the `planet traps' created by the surface density variations \citep{Masset2006}. Although we adopt a simple, proof-of-concept `toy model' for the generation of radial structuring of the disc, our results suggest that this may provide an effective means of allowing the formation of surviving cold Jupiters, and point the way to an avenue of potentially fruitful future research.

The paper is organised as follows.
We describe the updates to our physical model and numerical methods in sect. \ref{sec:methods}.
We present our results in sect. \ref{sec:results}, and  draw our conclusions in sect. \ref{sec:discussion}.

\section{Physical model and numerical methods}
\label{sec:methods}
The N-body simulations presented here were performed using the \emph{Mercury-6} symplectic integrator \citep{Chambers}, adapted to include the additional physical processes described below.
Some of these are updated versions of those described in \citet{ColemanNelson16}, and some of the processes are new to this paper.

\subsection{Recap of the physical model}
The disc and planet formation models from \citet{ColemanNelson16} contain the following elements:\\
(i) The N-body component computes the gravitational interaction between numerous planetary embryos and a swarm of orbiting planetesimals/boulders. The planetary embryos experience gravitational forces from each other and from the planetesimals/boulders, whereas the planetesimals/boulders experience gravitational forces from the embryos only, and not from the other small bodies. The planetesimals/boulders are treated as ``super-planetesimals'' that are supposed to represent a large collection of actual small bodies. The masses of these super-planetesimals are therefore much larger than the mass of an individual small body, but also significantly smaller than the embryo mass. The super-planetesimals experience gas drag, and are therefore assigned a physical radius in the range $10\,{\rm m} \le R \le 10 \, {\rm km}$. \\
(ii) The gas disc is evolved by solving the diffusion equation for a 1D viscous $\alpha$-disc model \citep{Shak,Lynden-BellPringle1974}.
Temperatures are calculated by balancing viscous heating and stellar irradiation with black-body cooling.
We use two values of the background $\alpha$ value in our simulations, $2\times 10^{-3}$ and $6 \times10^{-3}$. We assume that solar metallicity corresponds to a gas-to-solids ratio of 100:1, and
that half of the grains in the disc remain small and contribute to the opacity, and half of them grow to form larger planet-building bodies. When a giant planet is present, tidal torques from the planet are applied to the disc leading to the opening of a gap  \citep{LinPapaloizou86}.\\
(iii) We mimic the presence of fully-developed MHD turbulence in the inner disc, when the disc temperature exceeds 1000~K \citep{UmebayashiNakano1988, DeschTurner2015}, by increasing the viscosity parameter there by a factor of 5 above the background values of $\alpha=2 \times 10^{-3}$ or $6 \times 10^{-3}$.\\
(iv) We incorporate models for the photoevaporation of the disc. We have implemented a standard photoevaporative wind model \citep{Dullemond}, where the wind is assumed to be launched thermally from the disc upper and lower surfaces beyond a critical radius that corresponds approximately to the thermal velocity being equal to the escape velocity. We have also implemented a direct photoevaporation model that can be switched on, in addition to the standard model, if the inner gas disc becomes depleted and optically thin such that the stellar radiation acts directly on the disc inner edge that faces the star \citep{Alexander09}. The influence of including, or not including, the direct photoevaporation component on the final outcomes of the simulations is one of the issues that we investigate in this paper, along with the effect of neglecting photoevaporation altogether.\\
(v) Boulders and planetesimals orbiting in the disc experience aerodynamic drag, with the drag law (Stokes or Epstein) depending on the objects size and the molecular mean free path \citep{Weidenschilling_77}. For reference, we note that the drag-induced migration time scales associated with the different sized bodies are as follows (assuming a disc mass equivalent to the minimum mass solar nebula). A 10~m boulder migrates from $1 \au$ into the central magnetospheric cavity at $0.04\au$ in approximately $10^3$ yr, and from $20 \au$ in approximately $10^6$ yr. A 100~m planetesimal migrates from $1 \au$ within approximately $5 \times 10^5$ yr, and within the disc lifetime it migrates from $10 \au$ to $6 \au$. The 1~km and 10~km planetesimals display very small amounts of migration over the disc lifetime.
\\
(vi)  Torque formulae from \citet{pdk10,pdk11} are used to simulate type I migration due to Lindblad and corotation torques, and our model accounts for the possible saturation of the corotation torque. The influences of eccentricity and inclination on the disc forces are also included \citep{cressnels,Fendyke}.\\
(vii) Type II migration of gap forming planets is simulated using the impulse approximation of \citet{LinPapaloizou86}, where we use the gap opening criterion of \citet{Crida} to determine when to switch between type I and II migration. Thus, when a planet is in the gap opening regime, the planet exerts tidal torques on the disc to open a gap, and the disc back-reacts onto the planet to drive type II migration in a self-consistent manner.\\
(viii) The accretion of gaseous envelopes on to solid cores occurs once a planet's mass exceeds $3\me$, and the gas accretion rate is obtained from analytic fits to the detailed 1D models based on solving the planetary structure equations in the presence of prescribed rates of planetesimal accretion presented by \citet{Movs}. 
Gas that is accreted onto a planet is removed from the surrounding disc, such that the accretion scheme conserves mass.\\
(ix) The effective capture radius of a protoplanet that accretes boulders/planetesimals is enhanced by atmospheric drag using the prescription from \citet{Inaba}.

We adopt an inner boundary to our simulation domain at $0.04\au$, which we assume to represent the outer edge of an inner magnetospheric cavity. Any planet that enters this region no longer evolves, unless another planet enters the cavity, in which case the latter body is retained and the former one is assumed to have been pushed into the star. This is repeated for all subsequent planets that pass through the inner boundary (note that no sub-Neptune mass planets entered the cavity and pushed any giants into the star). When presenting our results in Figs. \ref{fig:massvperiod}, \ref{fig:CDFcomp}, \ref{fig:allCDFs} and \ref{fig:compsystems}, we reassign the final semimajor axes of these inner planets to straddle the stopping radius at $0.04\au$, in order to mimic our expectation that the inner cavities will have a range of radii. This reassignment assumes that the distribution of cavity edges is Gaussian with standard deviation of $0.01\au$.

\subsection{Model improvements}
\subsubsection{Gas envelope accretion}
\label{newgasaccretion}
A planet undergoes runaway gas accretion once the envelope and core are of comparable mass, and during this phase the planet rapidly accretes the material occupying its feeding zone, until it reaches its `gas isolation mass', where the feeding zone is now empty and a gap has formed in the disc.
\citet{ColemanNelson14} obtained fits to the runaway gas accretion rates from 2D hydrodynamic simulations, but they only considered the migration of the rapidly accreting planet after it had reached its `gas isolation mass'. We have improved on the fits of \citet{ColemanNelson14} by allowing the planet to migrate while undergoing runaway gas accretion. Including migration in the determination of the fits makes them more consistent with the hydrodynamic models.

Once a planet enters the runaway gas accretion phase prior to reaching the gap forming mass, we apply the following steps:\\
(i) Calculate the gas isolation mass, $m_{\rm iso}$, according to:
\begin{equation}
m_{\rm iso} = 2\pi r_{\rm p} \Sigma_g(r_{\rm p})\Delta r
\end{equation}
where $\Sigma_g (r_{\rm p})$ is the gas surface density taken at the planet's location, and $\Delta r$ is given by
\begin{equation}
\Delta r = 6\sqrt{3}R_{\rm H}
\end{equation}
where $R_{\rm H}$ is the planet's Hill radius.\\
(ii) Recalculate $m_{\rm iso}$ at each time step to account for the drop in $\Sigma_{\rm g}$ as the material in the planet's feeding zone diminishes.\\
(iii) Allow the planet to grow rapidly to $m_{\rm iso}$ by removing gas from the disc around the planet and adding it to the planet, using gas accretion rates obtained from the fits to the \cite{Movs} models.
Once the planet reaches $m_{\rm iso}$, it transitions to type II migration and begins accreting at the viscous rate.

When implementing the above prescription, we define the point at which the planet enters runaway gas accretion to be when $\frac{dm}{dt}\ge 2\me$ per 1000 yr. 
When the gas isolation mass is calculated we assume a maximum gas isolation mass of $400 \me$, which accounts for when a planet enters the runaway gas accretion phase in a massive disc, where tidal torques from the planet would evacuate the feeding zone before the gas isolation mass was reached.
We note that a planet that does not reach the runaway gas accretion mass prior to reaching the local gap forming mass would instead transition directly to type II migration without accreting the material within its feeding zone, and will begin accreting at the smaller of the rate obtained from the fits to \citet{Movs}, or the viscous supply rate.

\subsubsection{Migration during runaway gas accretion}
Until the planet reaches the mass required for runaway gas accretion, it undergoes type I migration using the torque formulae of \citet{pdk10,pdk11}.
Once it undergoes runaway gas accretion, the planet begins to carve a gap in the disc by rapidly accreting the surrounding material.
To account for this change in conditions, the planet stops undergoing type I migration and begins to migrate at a rate with a timescale equal to the local viscous evolution time:
\begin{equation}
\tau_{\nu}=\dfrac{2r_{\rm p}}{3\nu}.
\end{equation}
Migration at this rate continues until the planet reaches the gas isolation mass, where it transitions to self-consistent type II migration driven by the coupling to the viscous evolution of the disc via the impulse approximation \citep{LinPapaloizou86}. We note that recent hydrodynamic simulations have indicated that the migration of gap forming planets does not necessarily occur at exactly the viscous flow rate of the gas in the disc \citep{Duffell2014, DurmannKley2015}, due to residual gas in the gap adding to the migration torque. For the disc and planet masses that we consider in this work, however, the migration rates provided by the impulse approximation are in reasonable agreement with those obtained in hydrodynamic calculations \citep{ColemanNelson14}. 

\subsection{Disc radial structures}
\label{sec:radial}
Our recent simulations failed to form any surviving gas giant planets, other than hot Jupiters that are only prevented from migrating into their host stars by the presence of a central magnetospheric cavity \citep{ColemanNelson14, ColemanNelson16}. 
An analysis of the conditions required for gas giants to form and survive outside of the central cavity presented in \cite{ColemanNelson14} demonstrated that runaway gas accretion and the transition to type II migration needs to occur when the planets are distant from their stars. For example, for a Jovian mass planet to form and settle into a final orbit at $\sim 1\au$, requires type II migration to be initiated at $\sim 6\au$. A Jovian planet orbiting at $\sim 5 \au$ needs to initiate runaway gas accretion and type II migration at $\sim15 \au$. The time of formation also provides a constraint: form too early in the disc life time and a planet migrates all the way into the central cavity; form too late and there is insufficient gas available to build a gas giant. It is noteworthy that population synthesis simulations produce a large number of surviving cold gas giants \citep[e.g.][]{Mords09}.
\cite{ColemanNelson14} examined the planet mass and orbital evolution obtained using the following three approaches: 1D disc models similar to those presented in this paper; 2D hydrodynamic simulations that were designed to match the conditions in the 1D models; the prescriptions for mass growth and migration used in population synthesis models. They showed that the discrepancy obtained in giant planet survival rates between the modelling approaches arises because a migration-slowing factor is included in the population synthesis models when in the so-called planet-dominated regime, and this results in too much slowing of type II migration compared to that observed in the 2D hydrodynamic simulations or in the 1D viscous disc models.

Retaining the cores of gas giants at large orbital radii is difficult, especially late in the disc life time. The corotation and Lindblad torques need to balance, such that the core orbits in a ``zero-migration zone" \citep{BitschKley11,Hellary,Cossou}. The corotation torque has entropy-related and vortensity-related components, and it is the entropy-part that is normally strongest and able to balance the Lindblad torque when the temperature profile decreases outwards steeply. In a viscous, irradiated disc, the inner regions of the disc, where the viscous dissipation dominates the heating, have steep temperature gradients, and early in the disc life time the zero-migration zone can extend out to 
$\sim 10\au$ for planet masses $\gtrsim 10 \me$ \citep{BitschKley11,Hellary,Cossou,ColemanNelson14,Bitsch15a}. As the disc evolves, however, the viscous heating rate decreases and the zero-migration zone moves into the inner $1$--$2\au$ and only prevents the migration of lower mass planets. Although the details of the evolution depend on input parameters such as the viscous stress and the opacity, it would seem to be difficult to maintain strong entropy-related corotation torques in the outer disc regions during the later phases of disc evolution.
One alternative for maintaining cores at large radii might be for the vortensity-part of the corotation torque to be strengthened in regions where the surface density increases with radius,
such as may occur if the disc surface density contains undulations . These regions might act as planet traps \citep{Masset2006}, as well as being regions where small sized bodies such as dust, pebbles, boulders and small planetesimals could concentrate \citep[e.g.][]{Pinilla2012}. The main focus of this paper is to examine the consequences of allowing protoplanetary discs to be radially structured because of radial variations in the viscous stress. Our approach is to employ a very simple ``toy model" for simulating these radial structures, but we derive motivation from recent observations of protoplanetary discs, and from the long history of MHD simulations showing that discs which support magnetorotational turbulence \citep{BalbusHawley1991} often demonstrate radial structuring in the form of zonal flows.
  
\subsubsection{Observed structures}
Recent observations of the young class I T Tauri star, HL Tau, have shown the presence of a number of quasi-axisymmetric rings, corresponding to maxima and minima in the emitted intensity as a function of radius. The system of rings extends between 13--100$\au$ \citep{HLTAU}. A number of suggestions have been put forward to explain the rings, included embedded planets \citep{HLTAU_planets, HLTAU_KLEY},
pressure bumps that trap dust \citep{HLTAU_FLOCK}, enhanced dust growth near ice lines \citep{HLTAU_ZHANG}, and sintering of dust aggregates \citep{Okuzumi2015}.
Even more recent ALMA observations of the disc around TW Hydra have also uncovered a series of rings \citep{Andrews2016}, suggesting that these really are common phenomena that arise during the evolution of protoplanetary discs. The closer proximity of TW Hydra to the Solar System allows regions of the disc that lie closer to the central star to be probed by the ALMA observations, and these have uncovered rings between orbital radii 1 -- $40 \au$. Furthermore, although high resolution ALMA images of other protoplanetary discs have not yet been released, existing ALMA data for a number of other discs indicate that ring structures are present in the disc outer regions \citep{Zhang2016}, suggesting that these features are common phenomena that arise during the evolution of protoplanetary discs. Although we do not attempt to fit our simulations to these observations, we simply note that a plausible scenario for the origin of these rings is radial variation in the effective viscous/turbulent stresses that give rise to variations in the surface density.

\begin{figure}
\includegraphics[scale=0.45]{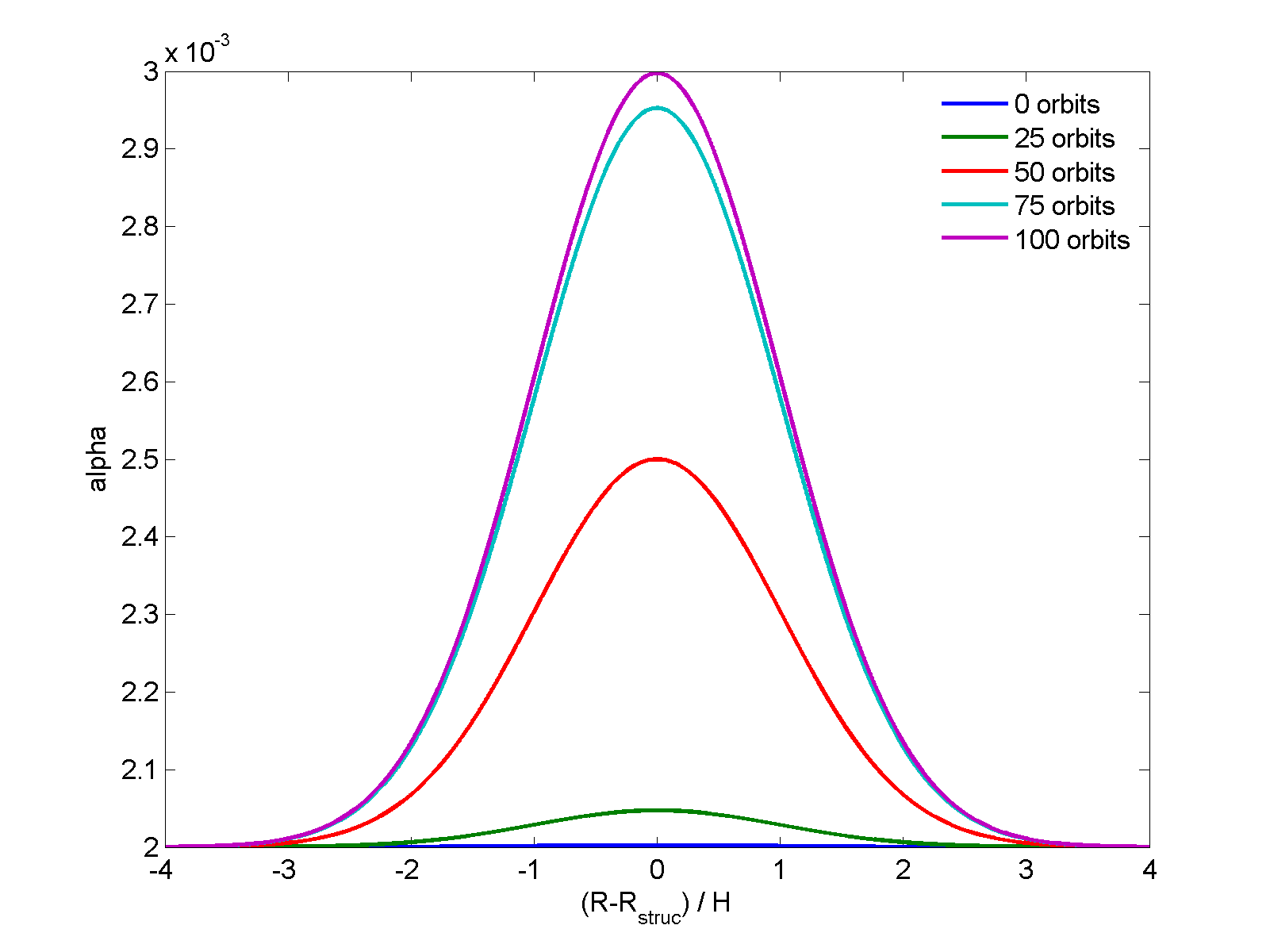}
\caption{Plot showing the time variation of the viscous $\alpha$ associated with the formation of a radial structure over a time of 100 local orbits.}
\label{fig:intialzonalflow}
\end{figure}

\begin{figure*}
\includegraphics[scale=0.4]{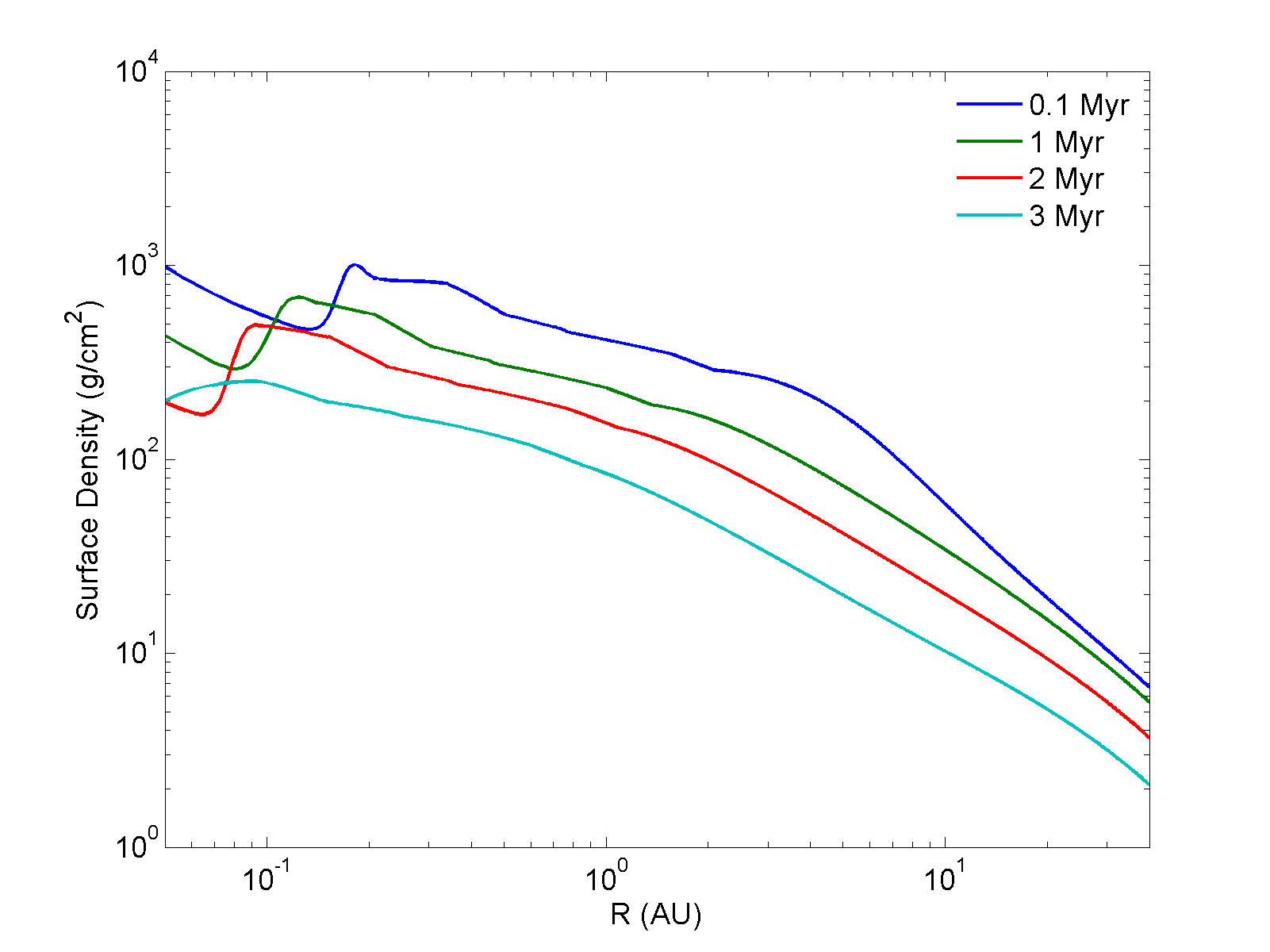}
\includegraphics[scale=0.4]{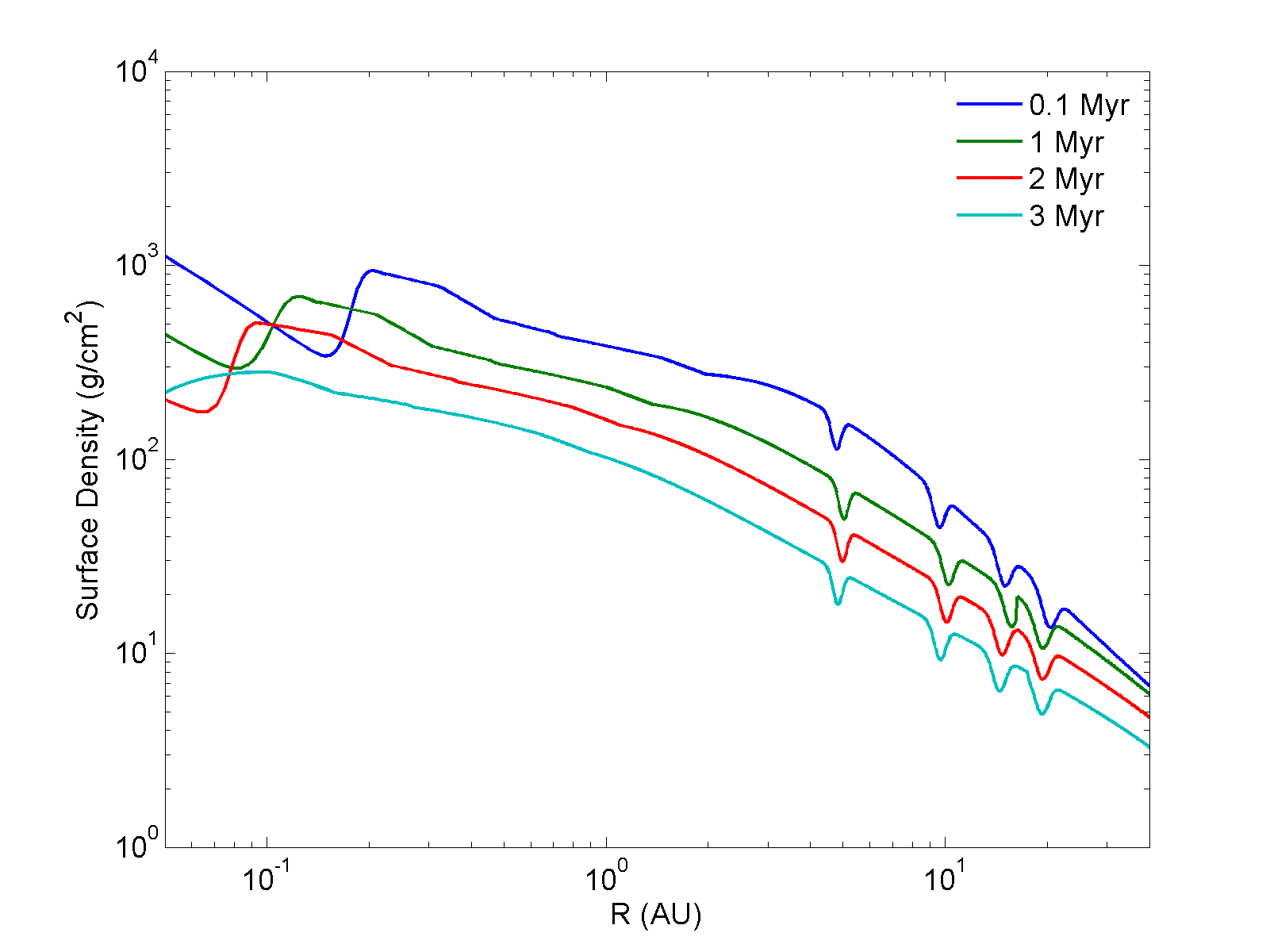}
\caption{Surface density profiles at t = 0.1, 1, 2, 3 Myr for a $1\times \mmsn$ disc (total lifetime $\sim 5.5$ Myr) without (left panel) and with (right panel) radial structuring.}
\label{fig:sigma_comp}
\end{figure*}

\subsubsection{Zonal flows in MHD simulations}
Over a number of years, both global \citep{SteinackerPapaloizou2002, PapaloizouNelson2003, FromangNelson2006} and local \citep{Johansen2009} simulations of magnetised discs have demonstrated the occurrence of persistent density/pressure maxima and minima as a function of radius, arising from localised magnetic flux concentration and associated enhancement of magnetic stresses. More recent simulations incorporating non-ideal MHD effects have also reported the existence of these features in local \citep{Bai2014}
and global \citep{ZhuStoneBai2014,BethuneLesur2016} simulations. Density variations with amplitudes up to $\sim 50 \%$ of the background have been reported \citep{Bai2014}. Being in geostrophic balance, these pressure bumps are often referred to as zonal flows \citep{Johansen2009}. \cite{Simon12} have recently observed long-lived zonal flows in simulations with radial domains up to 16 scale heights, and in these large shearing boxes they find that the outer radial scale of the zonal flows is  $\sim 6H$, although they stress that simulations in larger domains are required to demonstrate convergence. \citet{Dittrich13} ran shearing box  simulations with radial domains up to $21H$ and also found the radial sizes of the axisymmetric zonal flows to be between 5 and $7H$. The study by \citep{Bai2014} noted that zonal flows in radially-narrow shearing boxes tended to be intermittent, but runs in large shearing boxes of width $16 H$ persisted for the full duration of the simulations, which had total run times of 400 orbits. Zonal flows are clearly able to live for long times, but at present it is not clear what their characteristic life times are.

\begin{table}
\centering
\begin{tabular}{cccc}
\hline
Structure & $R_{\rm min}$ & $R_{\rm max}$ & Lifetimes\\
 label & (au) & (au) & ($\times10^3$ local orbits)\\
\hline
1 & 4.25 & 5.75 & 10, 50, 100\\
2 & 9.25 & 10.75 & 10, 50, 100\\
3 & 14.25 & 15.75 & 10, 50, 100\\
4 & 19.25 & 20.75 & 10, 50, 100\\
\hline
\end{tabular}
\caption{Radial structure parameters}
\label{tab:zonal}
\end{table}

Although we do not try to fit a model to these MHD simulations, and instead take the approach of employing a simple prescription to demonstrate ``proof of concept", we note that global MHD simulations which display dust concentration in pressure bumps have been used to compare theoretical calculations with the observed structures in protoplanetary discs \citep{HLTAU_FLOCK}.
Density and pressure bumps arising from variations in magnetic or turbulent stresses may be a common feature of planet forming discs. In addition to the zonal flows described above, similar features may also arise in regions where there is a transition from one non-ideal MHD process being dominant to another becoming dominant (e.g. a transition between Hall and ambipolar dominated regimes), or at the interface between magnetically active and dead zones. In the presence of these transitions, the disc may not be able to maintain a constant mass flux through all radii at all times, and radial structuring may occur. For simplicity, in this paper we just consider a rather crude model for disc structuring that is intended to mimic the growth and decay of zonal flows, but we note that radial structuring may also occur because of other physical processes that influence the local rate of mass flow through the disc.

\subsubsection{A simple model for radial structuring}
We incorporate radial structuring in our models by introducing a spatially and temporally varying viscous stress. At any one time, four structures are present in our simulations. While this number is arbitrary, it is similar to the number of rings observed in HL Tau and TW Hydra. Each one exists between specific, predefined radii ($R_{\rm min}$, $R_{\rm max}$), where the values are given in Table \ref{tab:zonal}. Each structure has a finite life time (see the final column of Table \ref{tab:zonal}), and as it decays a new structure grows within the same range of radii $R_{\rm min} <  R <  R_{\rm max}$. We initiate the structures 50,000 years after the start of the simulations, once the disc has reached a quasi-steady state, by increasing the viscosity parameter $\alpha$ up to a maximum strength of $1.5 \times$ that of the background value. This value was chosen to approximately match the $\sim 50 \%$ variation in the surface density due to the zonal flows obtained in the MHD simulations of \citet{Bai2014}. For each structure, the maximum value of $\alpha$ is located at the centre of that structure, whilst we transition $\alpha$ to its background value over a distance of 3.5 local scale heights using a Gaussian kernel, giving each structure a width of 7 local scale heights. Once each of the structures begins to form, it does so over 100 local orbital periods by increasing $\alpha$ from the background value up to the required value, as described below:
\begin{equation}
\label{eq:alphanew}
\alpha(r,t ) = \alpha_{\rm b} + \dfrac{\alpha_{\rm b}}{2} R_{\rm new} t_{\rm new}
\end{equation}
where $\alpha_{\rm b}$ is the background value, and $R_{\rm new}$ and $t_{\rm new}$ are the radial and time factors defined by
\begin{equation}
\label{eq:rnew}
R_{\rm new} = \exp{\left(\dfrac{-(R-R_{\rm struc})^2}{2H^2_{\rm struc}}\right)} 
\end{equation}
\begin{equation}
t_{\rm new} = 0.5\times\left(\tanh{\left(\dfrac{6(t-t_{\rm start}-0.5(t_{100}-t_{\rm start}))}{t_{100}-t_{\rm start}}\right)}+1\right)
\end{equation}
where the subscript `struc' denotes the radial location of the centre of the structure, $t_{\rm start}$ is the structure formation time, $t_{100}-t_{\rm start}$ represents the time interval of 100 orbital periods after the structure begins to form, evaluated at the structure's centre, and $H$ is the local disc scale height.
The shape and time evolution of the locally varying viscous $\alpha$ parameter associated with an individual structure as it forms is shown in Fig. \ref{fig:intialzonalflow}, where the $\alpha$ parameter gradually increases to the required value, while maintaining a smooth profile.

Our structures have specific lifetimes, and this is a parameter that we vary in the simulations described below. When a structure comes to the end of its lifetime it quickly disappears over 100 local orbital periods. As one structure disappears, another one forms at a randomly chosen location within the range
of allowed radii given in Table \ref{tab:zonal}. When the structure starts to disappear, $\alpha$ evolves according to:
\begin{equation}
\alpha(r) = \alpha_{\rm b} + \dfrac{\alpha_{\rm b}}{2} R_{\rm old} (1-t_{\rm old}),
\end{equation}
where $t_{\rm old}$ is given by
\begin{equation}
t_{\rm old} = 0.5\times\left(\tanh{\left(\dfrac{6(t-t_{\rm end}-0.5(t_{100}-t_{\rm end}))}{t_{100}-t_{\rm end}}\right)}+1\right).
\end{equation}
Here, $t_{\rm end}$ is the time at which the structure begins to dissipate and $t_{100}$ represents 100 orbital periods after this time.
$R_{\rm old}$ is equal to equation \ref{eq:rnew} but with values taken for the old structure instead of a new one, as shown by this expression
\begin{equation}
R_{\rm old} = \exp{\left(\dfrac{-(R-R_{\rm oldstruc})^2}{2H^2_{\rm oldstruc}}\right)}. 
\end{equation}
To account for a new structure being influenced by a dissipating older structure, equation \ref{eq:alphanew} becomes
\begin{equation}
\label{eq:alphanewnew}
\alpha(r, t) = \alpha_{\rm b} + \dfrac{\alpha_{\rm b}}{2}(R_{\rm new}t_{\rm new}+R_{\rm old}(1-t_{\rm old}))
\end{equation}
This allows a smooth transition between two adjacent forming/dissipating structures.

Below we discuss the main effects of radial structures on the disc profile and migration of embedded low mass planets.

\subsubsection{Effects on disc and planet evolution}
\label{sec:effect}

\begin{figure*}
\includegraphics[scale=0.43]{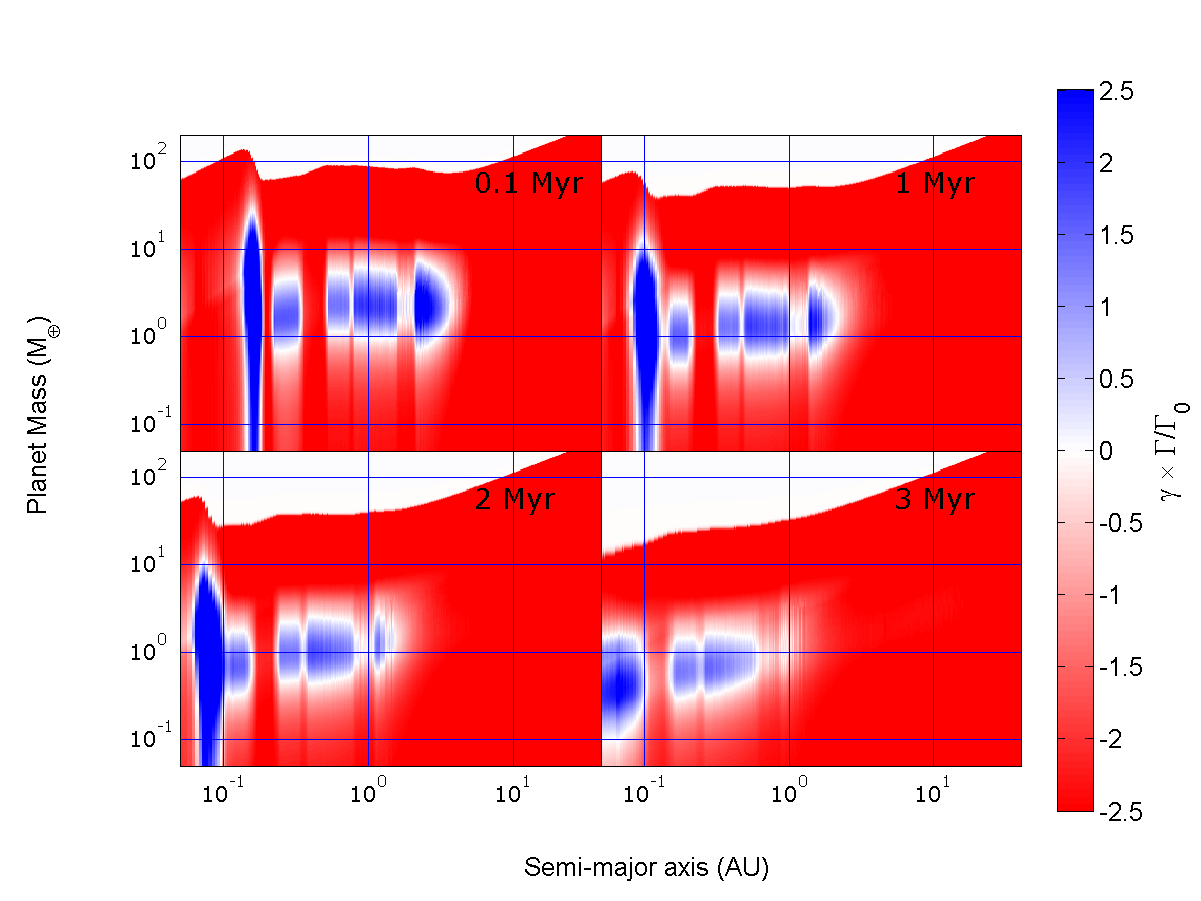}
\includegraphics[scale=0.43]{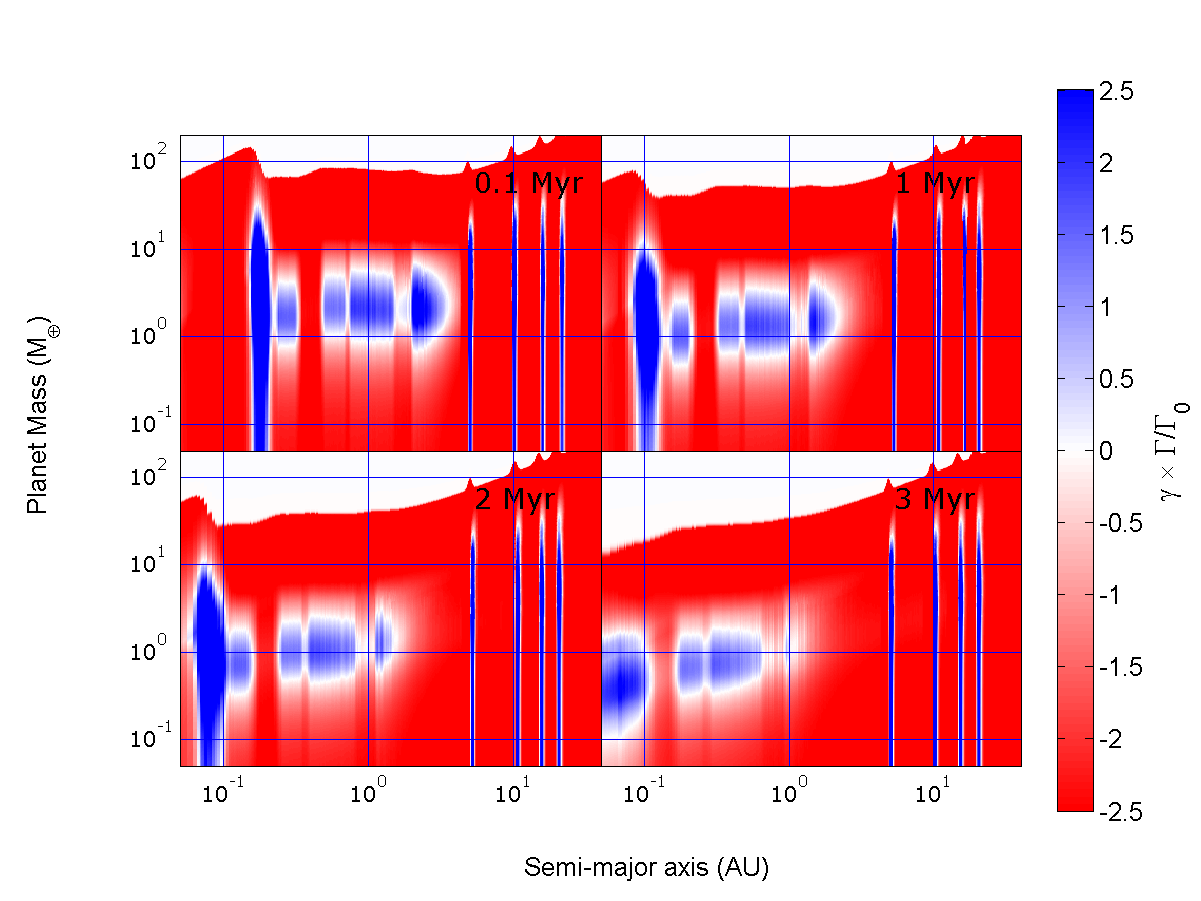}
\caption{Contour plots showing regions of inwards (red) and outwards (blue) migration) in a $1 \times \mmsn$ disc at t = 0.1 (top left), 1 (top right), 2 (bottom left) and  3 Myr (bottom right) for discs without (left panel) and with radial structuring (right panel). The white contours at the top of each panel corresponds to the planet reaching the local gap forming mass, at which point the planet will undergo type II migration. The contours represent values of $\gamma \Gamma/\Gamma_0$, where $\gamma$ is the ratio of specific heats, $\Gamma$ is the torque experienced by a planet and $\Gamma_0$ is a normalisation factor defined in \citet{pdk10}}
\label{fig:contour_comp}
\end{figure*}

Fig. \ref{fig:sigma_comp} shows the surface density evolution for a $1\times \mmsn$ disc in simulations without (left panel) and with (right panel) radial structuring. The drop in surface density in the inner regions of the discs arises because both models include an increase in $\alpha$ by a factor of 5 from the background values of either $2 \times 10^{-3}$ or $6 \times 10^{-3}$ where the disc temperature exceeds 1000~K \citep{ColemanNelson16}. The presence of the radial structures arising from the variations in $\alpha$ in the outer disc are evident in the right panel. While these surface density dips have little influence on the global disc evolution, they have a dramatic effect on planet migration.

Fig. \ref{fig:contour_comp} shows contours that illustrate the direction and speed of type I planet migration as a function of planet mass and semimajor axis at different times in a  
$1 \times \mmsn$ disc with solar metallicity.
The left panel shows a simulation without radial structuring, and the right panel is for a run with structuring switched on. Red regions correspond to rapid inward migration, blue regions correspond to rapid outward migration, and white contours interspersed between the red and blue contours represent zero-migration zones where Lindblad and corotation torques cancel. The white contours at the top of the panels correspond to planets reaching the gap opening mass and undergoing type II migration. The planet trap arising from the inner fully-developed turbulent region is represented by the innermost blue contour, apparent in the first three frames of each simulation. As the surface density decreases, the zero-migration zones and extended regions of outward migration associated with strong entropy-related corotation torques slowly move in towards the star on the disc evolution time scale in both runs, but the run with radial structures maintains four zero-migration zones in the outer disc for the duration of the simulation, leading to the possibility of long-term trapping of planetary cores with masses up to $\sim 30 \me$.

If a planet core was to migrate to the edge of one of the four structures, then it would be trapped for the lifetime of the structure. The core is released from the structure when it comes to the end of its life, and the planet starts to migrate inwards. A new structure is formed locally to replace the old one, and this has some probability of being located inside the old one (that depends on the location of the old structure within its allowed range of radii). If the new structure sits inside the old one then the planet core can be trapped by it, but if it sits outside the planet location then the planet migrates inwards, either into one of the other three structures, or in towards the star if it has just escaped from the innermost structure. Furthermore, a rapidly migrating planet core can escape from a structure while it is decaying and before the next structure has developed fully. This shows that the long term orbital evolution of a planetary core has a stochastic element that depends on the detailed evolutionary histories of the radial structures in the disc. Some cores remain trapped at large radius over the disc lifetime, whereas other cores escape from the planet traps and migrate into the disc inner regions.

Gas accretion can occur onto a core that is trapped if its mass exceeds $m_{\rm p}\ge3\me$, and if it remains in the outer disc for an extended period of time then runaway gas accretion can occur and a giant planet can form. The planet would then open a gap in the disc, and begin to undergo type II migration as the planet traps are not effective for gap forming planets. The process of building planets at the planet traps is enhanced by the concentration of boulders and planetesimals at these locations, which can then be accreted efficiently by the growing planets. In general, we find that accretion of solids by planetary embryos occurs during an early burst, prior to the onset of the main gas accretion phase. Gas accretion is then accompanied by modest planetesimal accretion at rates that are similar to or below those prescribed in the \citet{Movs} models that determine gas accretion rates in our simulations. Approximately 20\% of the giant planets in our runs experience an episode of rapid and short-lived solids accretion, normally during the runaway gas accretion phase when the growth of the giant acts to destabilise the system. This burst of accretion can either arise from an impact with a low mass protoplanet, or through accretion of a local swarm of planetesimals over a time period that is less than $\sim$10,000 years. 
Our fits to the \cite{Movs} models do not allow the gas accretion rate to respond to this time-varying planetesimal accretion, and this is one area for future improvement of our model.

In summary, we have introduced a simple model for the radial structuring of protoplanetary discs that includes assumptions about the number of surface density features (planet traps) that are formed and their lifetimes. We present this model as a simple proof-of-concept in this paper, and do not include an extensive analysis of what happens when the model parameters are modified. It is reasonable to suppose, however, that reducing the number of planet traps and their lifetimes will result in less efficient trapping of planet cores, and hence less efficacious giant planet formation. Precisely how the formation of giant planets is affected by variation of model parameters will be examined in future work.

\subsection{Initial conditions}
\label{sec:initial}
Table \ref{tab:parameters} gives an overview of the parameters used in our simulations.
All simulations were initiated with 44 planetary embryos, of mass $0.2\me$, with semimajor axes between 1 and 20 $\au$ and separated by 10 mutual Hill radii.
These were embedded in a swarm of thousands of planetesimals/boulders, that were distributed with semimajor axes between 0.5 and 25 $\au$, with masses either 10, 20 or 50 times smaller than the embryos, depending on the metallicity of the system. (This varying mass ratio between embryos and planetesimals was implemented to obtain a planetesimal number that allowed the simulations to run on reasonable time scales. Between 1000 and 5000 planetesimals/boulders were used and run times for individual simulations varied between 2 and 6 months.)
The total mass of solids ranges between 12.5--109$\me$ depending on the disc mass and metallicity.
The effective physical radii of planetesimals were set to 10~m, 100~m, 1~km or 10~km, such that the primary feedstock of the accreting protoplanets ranged from being boulders to large planetesimals whose evolution differed principally because of the strengths of the gas drag forces that they experienced. 
Initial eccentricities and inclinations for protoplanets and planetesimals/boulders were randomized according to a Rayleigh distribution, with scale parameters $e_0=0.01$ and $i_0=0.25^{\circ}$, respectively.

Collisions between protoplanets and other protoplanets or planetesimals resulted in perfect sticking, which probably results in a slight overestimate of accretion rates in the simulations.
We neglect planetesimal-planetesimal interactions and collisions in our simulations for reasons of computational speed.

The gas disc masses simulated were 1 and 2 times the mass of the minimum mass solar nebula \citep[MMSN][]{Hayashi}. We also vary the metallicity so that the initial solids-to-gas mass ratios in the discs are equal to 0.5, 1 and 2 times the solar value for the different models. We define the solar metallicity to be equivalent to the solids-to-gas ratio introduced by \citet{Hayashi}. We smoothly increase the mass of solids exterior to the snow line by a factor of 4 by increasing the numbers of planetesimals, and the initial surface density of solids follows the initial gas surface density power law, as described in \citet{Hellary}. The factor of four increase in solids implies a larger increase in water content than obtained by \citet{Lodders2003} for the solar nebula, but is consistent with the widely used Hayashi model \citep{Hayashi}. We adopt this model to maintain continuity with our earlier work \citep{Hellary,ColemanNelson14,ColemanNelson16}. We track the changes in planetary compositions throughout the simulations, as planets can accrete material originating either interior or exterior to the snow line.

\begin{table}
\centering
\begin{tabular}{cc}
\hline
Parameter & Values/Ranges\\
\hline
Disc mass & 1, 2 $\times\mmsn$\\
Disc metallicity & 0.5, 1, 2 $\times$ Solar\\
Total solids mass & 12.5--109 $\me$\\
Background viscous $\alpha$ & $2 \times 10^{-3}$, $6\times10^{-3}$\\
Planetesimal radii & 10\,m, 100\,m, 1\,km, 10~km\\
Planetesimal mass & 0.004, 0.01, 0.02 $\me$\\
Planetesimal number & 1000 -- 5000\\
Gas disc lifetimes &3.5 -- 8.4~Myr \\
\hline
\end{tabular}
\caption{Values, and the ranges of values, adopted for various simulation parameters.}
\label{tab:parameters}
\end{table}

We use two different values for the background $\alpha$ value, $\alpha=2\times 10^{-3}$ and
$6\times 10^{-3}$.
These values of $\alpha$ correspond to disc lifetimes of 5.5 and 3.5 Myr respectively for a disc with mass equal to $1\times\mmsn$.
We examine the effect of varying the lifetimes of the radial structures in the disc, with the three values assumed being $10^4$, $5\times 10^4$ and $10^5$ local orbital periods.
We ran two instances of each parameter set, where only the random number seed used to generate initial particle positions and velocities was changed, meaning that a total of 288 simulations have been run.
The simulations were run for 10 Myr, or until no protoplanets remained.

\begin{table*}
\centering
\begin{tabular}{lccccc}
\hline
Classification & Mass & Rock $\%$ & Ice $\%$ & Gas $\%$ & Final Number\\
\hline
Rocky Earth & $m_{\rm p}<3 \me$ & $>70$ & $<30$ & $0$ & $1563$ \\
Water-rich Earth & $m_{\rm p}<3 \me$ & $<70$ & $>30$ & $0$ & $4625$ \\
Rocky super-Earth & $3 \me\leq m_{\rm p}<10 \me$ & $>60$ & $<30$ & $<10$ & $12$ \\
Water-rich super-Earth & $3 \me\leq m_{\rm p}<10 \me$ & N/A & $>30$ & $<10$ & $83$ \\
Gas-rich super-Earth & $3 \me\leq m_{\rm p}<10 \me$ & N/A & N/A & $>10$ & $69$ \\
Gas-poor Neptune & $10 \me\leq m_{\rm p}<35 \me$ & N/A & N/A & $<10$ & $5$ \\
Gas-rich Neptune & $10 \me\leq m_{\rm p}<35 \me$ & N/A & N/A & $>10$ & $79$ \\
Gas-poor super-Neptune & $35 \me\leq m_{\rm p}<100 \me$ & N/A & N/A & $<50$ & $29$ \\
Gas-rich super-Neptune & $35 \me\leq m_{\rm p}<100 \me$ & N/A & N/A & $>50$ & $147$ \\
Jupiter & $100 \me\leq m_{\rm p}<1000 \me$ & N/A & N/A & $>50$ & $120$ \\
Super-Jupiter & $m_{\rm p}\geq 1000 \me$ & N/A & N/A & $>50$ & $12$ \\
\hline
\end{tabular}
\caption{Planetary classification parameters based on their mass and composition. 
Note that water-rich planets are so-called because they accrete water ice in solid form that originates from beyond the snow-line. Characteristics that play no role in the classification of a planet are denoted by ``N/A" in the relevant columns. Note all Jupiters and Super-Jupiters formed in the simulations had gas mass fractions $\geq50\%$.}
\label{tab:classifications}
\end{table*}

\subsection{Planet classification scheme}

To assist in the discussion of simulation outcomes, we have developed a classification system for the different bodies that are formed. As there are no formal IAU definitions for exoplanet classes relating to their masses and compositions, there is freedom of choice in how planets should be classified. We have chosen a scheme that uses mass as the primary discriminant and composition as a secondary one. We use the labels ``Earth", ``Neptune" and ``Jupiter", along with the prefix ``super" to define six mass-based classes, and subclasses are defined according to the volatiles content, either in the form of ice or gas, of the planets.  Definitions of the different planet classes are given in Table \ref{tab:classifications}. Note that when we use the term ``gas giant" we are referring to Jupiters or super-Jupiters.

\section{Results}
\label{sec:results}

We now present the results for our simulations. We begin by discussing a representative run in which multiple giant planets were able to form and survive. We then present an overview of all the simulation outcomes, before examining how modifying parameters such as the disc mass, metallicity, photoevaporation model etc. changes the results.

\begin{figure*}
\includegraphics[scale=0.8]{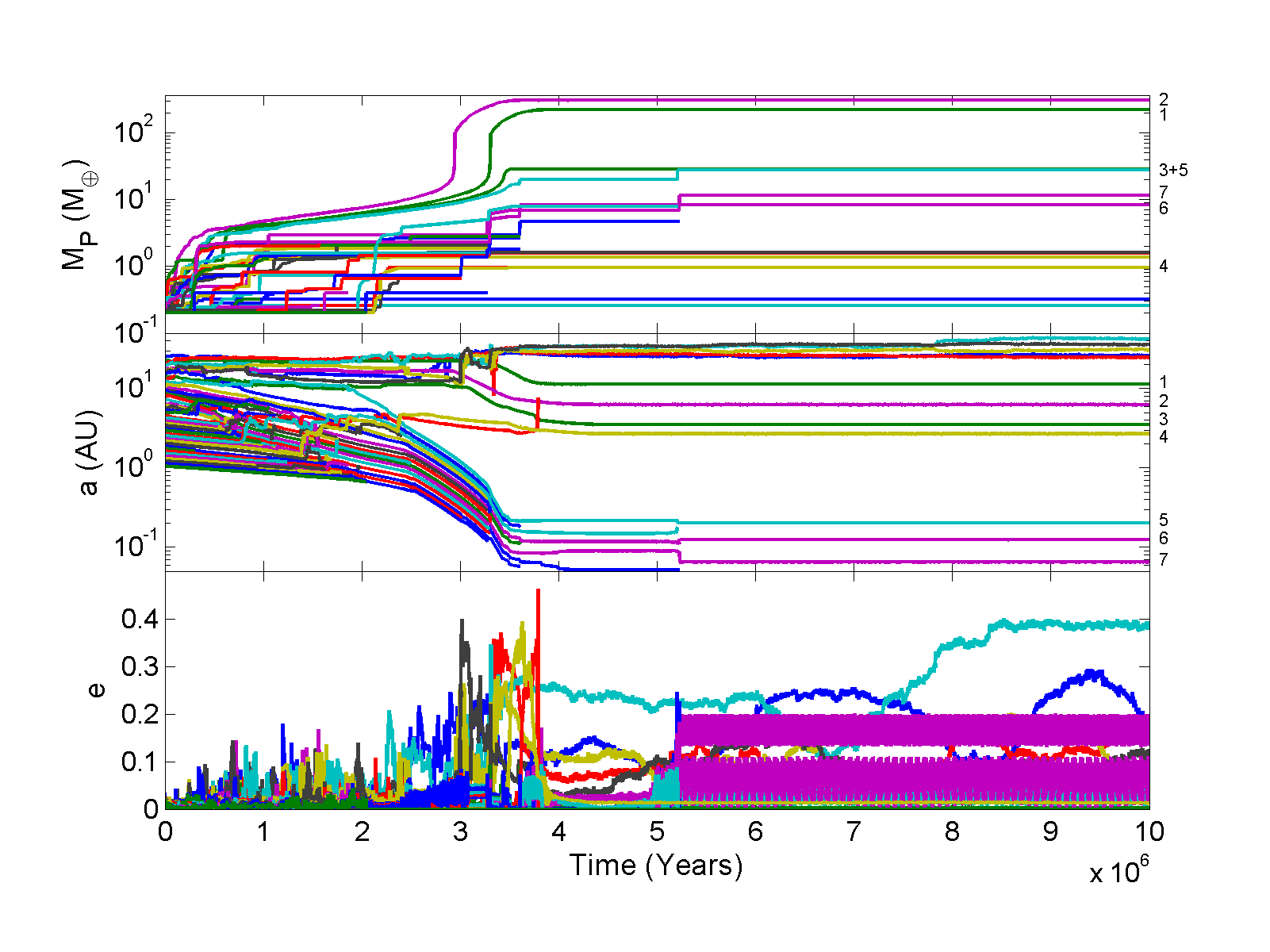}
\caption{Evolution of masses, semimajor axes and eccentricities of all protoplanets 
in simulation CJ120.1210A, where the disc lifetime $\sim 4.5$~Myr.
Note that formation histories of selected surviving planets are indicated by the labels on the right side of the mass and semi-major axis subplots.}
\label{fig:G120.1210Amulti}
\end{figure*}

\begin{figure}
\includegraphics[scale=0.45]{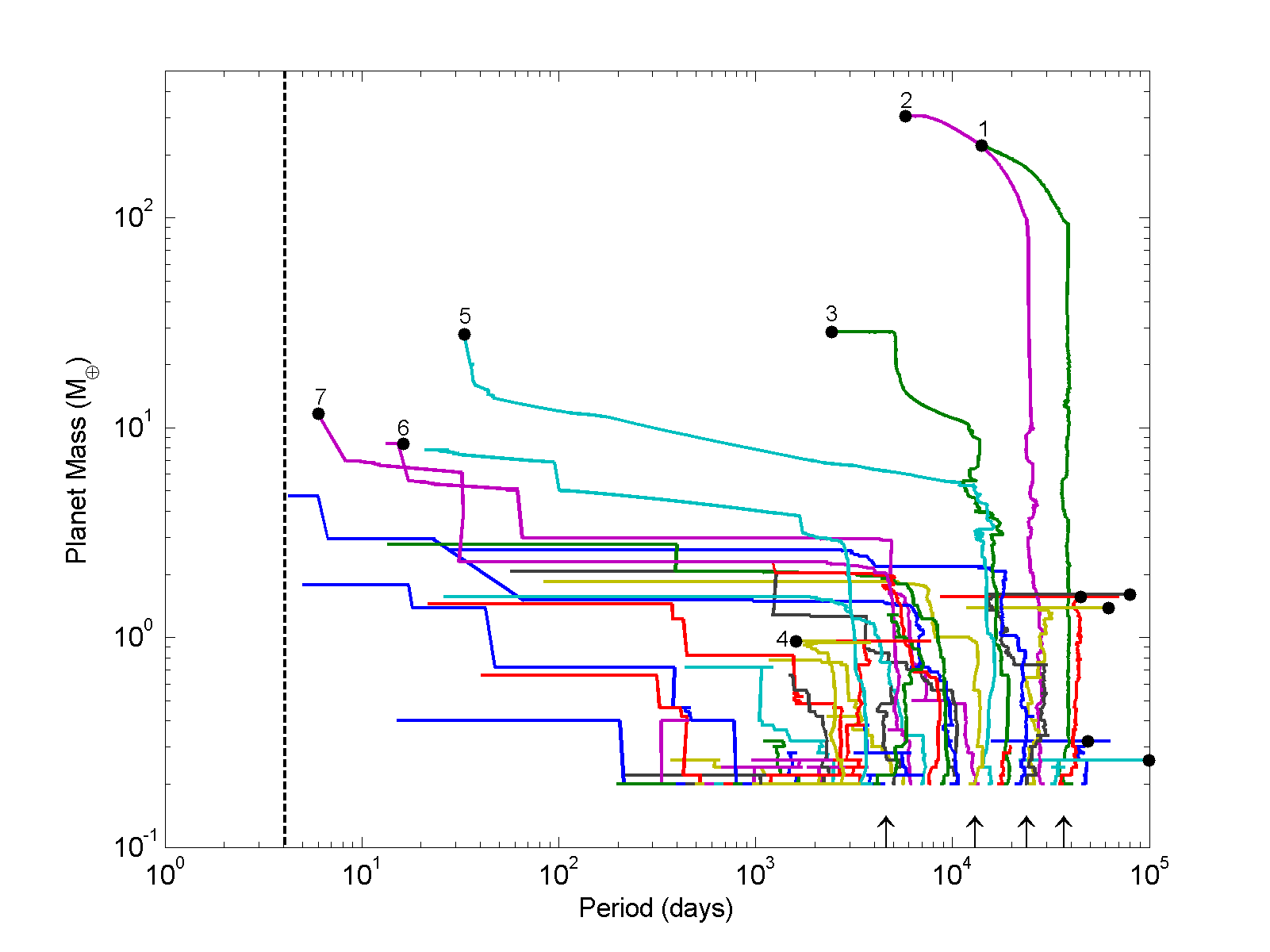}
\caption{Evolution of planet mass versus orbital period for all protoplanets in simulation CJ120.1210A.
Filled black circles represent final masses and orbital periods for surviving planets.
Note that formation histories of selected surviving planets are indicated by the labels adjacent to the filled black circles.
The dotted black line at $\sim 4$d represents the inner edge of our simulated protoplanetary disc.
The arrows above the x-axis indicate the average positions of the four radial structures.}
\label{fig:G120.1210AMVA}
\end{figure}

\subsection{Run CJ120.1210A}
\label{sec:example}
Run CJ120.1210A had a disc mass of $1\times\mmsn$, $2\times$ solar metallicity, and contained planetesimals with radii $R_{\rm p} = 100$~m. The total mass in planetesimals was $43.2\me$ and that in protoplanets was $8.8\me$. The background $\alpha = 2\times  10^{-3}$, and radial structures had a lifetime of 10,000 local orbits. The direct photoevaporation model was used.

The evolution of protoplanet masses, semimajor axes and eccentricities are shown in Fig. \ref{fig:G120.1210Amulti}, and the final state of the system is also represented in Fig.~\ref{fig:compsystems} (the case with the label CJ120.1210A).
We also show the mass versus orbital period evolution of all protoplanets in Fig. \ref{fig:G120.1210AMVA}, where filled black circles represent surviving planets, and the evolution of the labelled planets is described below.
The end state after 10~Myr consists of: an inner compact system comprising 3 super-Earths/Neptunes; a cool Neptune and Earth-mass planet orbiting between 2.5--$3.4\au$; two cold Jupiters orbiting between 6--$12\au$; a collection of low mass planets (`debris'), that failed to grow during the simulation,  orbiting out beyond 20-$30\au$. We ignore the long period `debris' in our discussion below, and just concentrate on the other planets that form.

\subsubsection{Cold Jupiters}
The cores of the two Jupiters (see planet labels 1 and 2 in Figs. \ref{fig:G120.1210Amulti} and \ref{fig:G120.1210AMVA}) begin to form at orbital radii 15-$25\au$ within the first 0.5~Myr of the simulation, through a combination of planetesimal accretion and mutual collisions between embryos. Migration and trapping of planetesimals in the radial structures helps concentrate material which is then accreted by the embryos, stimulating rapid growth above $3 \me$ such that gas accretion onto the growing cores can start. These proto-giant planets remain trapped at large radii by the radial structures, and continue to accrete gas steadily until runaway gas accretion is initiated at times just before and after 3~Myr, respectively (see the top panel of Fig.~\ref{fig:G120.1210Amulti}). The rapid burst of gas accretion takes the planet masses up to $\sim 100\me$, after which gap opening ensues. Initially both planets accrete at the viscous supply rate, but `planet 1' truncates the disc exterior to it and prevents further gas accretion on to `planet 2', which lies interior to `planet 1'. The onset of gap formation allows the planets to type II migrate inwards until the gas disc is completely removed after $\sim 4.5$~Myr, although we note that the migration of `planet 2' is slowed by the truncation of the disc by `planet 1'.
The gas giants have masses $306 \me$ and $222 \me$, gas mass fractions of 98\%, semimajor axes $6.3\au$ and $11.4\au$, orbital periods 15.8~yr and 38.5~yr and eccentricities $\sim0$, respectively. While this pair of planets are far from being a perfect analogue to the Jupiter-Saturn system, there is an obvious similarity in terms of gross characteristics that is worthy of note (i.e. mass of inner planet > mass of outer planet and semimajor axis ratio $\sim 1.8$).

\subsubsection{Cool Neptune and Earth}
These planets are labelled as 3 and 4 in Figs. \ref{fig:G120.1210Amulti} and \ref{fig:G120.1210AMVA}.
The cool Neptune begins its formation out beyond $10\au$ at the same time as the giant planet cores are forming, but interior to these two proto-giants. It also begins to accrete gas within the first 0.5~Myr, but at a slightly slower rate than the two proto-giants, and remains trapped by the radial structures during the first 3~Myr. The cool Neptune is nudged inwards when the innermost gas giant undergoes runaway gas accretion and starts type II migrating, and this allows the Neptune to escape the radial structures and migrate in towards the central star. Gas accretion onto the Neptune and its migration halt when the gas disc disperses after 4.5~Myr, leaving it with a mass of $28.6 \me$, gas mass fraction of 86\%, semimajor axis $\sim 3.5\au$, orbital period 6.5~yr, and eccentricity $\sim0$. As this gas-rich Neptune escaped the radial structures and migrated inwards it shepherded a $1\me$ water-rich terrestrial planet ahead of it, which had a final semimajor axis $\sim 2.7\au$ and orbital period $\sim$ 4.4 yr.

\subsubsection{Compact inner system of super-Earths/Neptunes}
The planets we discuss here have labels 5-7 in Figs. \ref{fig:G120.1210Amulti} and \ref{fig:G120.1210AMVA}.
This compact system forms from a combination of bodies that are initially orbiting interior to the radial structures and one dominant body that originates from larger radii. This more massive body grows through planetesimal accretion and collisions with neighbouring embryos out beyond $10\au$, where it starts to accrete gas and remains trapped by the radial structures until 2~Myr. At this point its mass is $5 \me$, and it is able to escape from the radial structures by migrating through them as they switch on and off, after which it undergoes rapid inward type I migration while continuing to accrete gas (becoming a gas-rich Neptune in the process). The gas-rich Neptune shepherds a large number of interior embryos in a resonant convoy as it migrates, and when the gas disc starts to disperse after $\sim 3.5$~Myr this convoy breaks up and mutual collisions between the numerous embryos lead eventually to the formation of a compact inner system comprised of 3 planets: a gas-poor Neptune with mass $11.6\me$, gas mass fraction 7\%, semimajor axis $\sim 0.07\au$, orbital period 6 days and eccentricity 0.11; an icy super-Earth with mass $8.4 \me$, gas mass fraction 6.5\%, semimajor axis $\sim 0.15\au$, orbital period 16.2 days and eccentricity 0.2; a gas-rich Neptune with mass $27.6\me$, gas mass fraction 56\%, semimajor axis $\sim 0.2\au$, orbital period 33.2 days and eccentricity 0.07. We note that the eccentricities of the these planets were pumped up to the values shown in the bottom panel of Fig.~\ref{fig:G120.1210Amulti} during a late scattering event at 5.2 Myr.

\begin{figure*}
\includegraphics[scale=0.4]{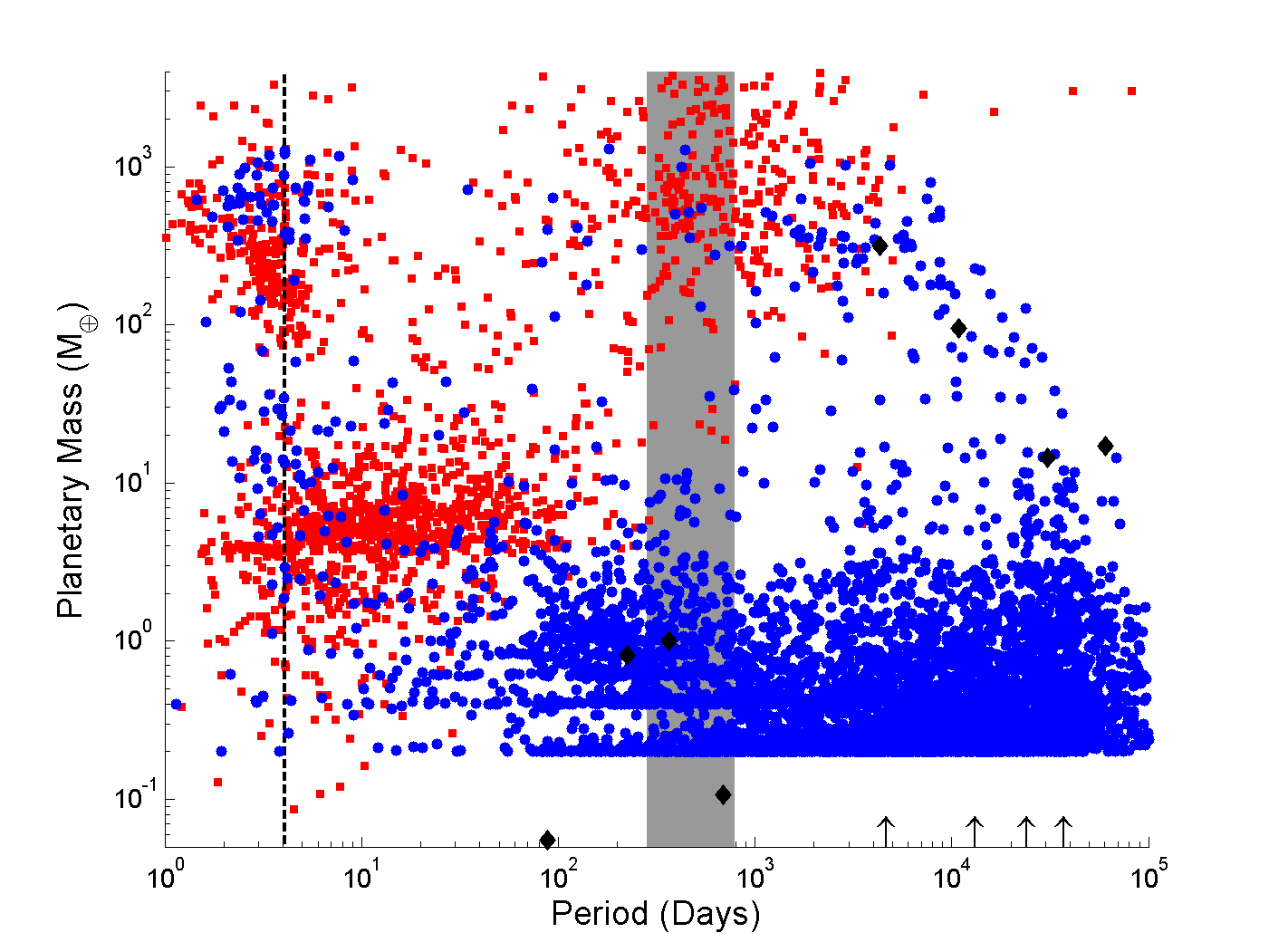}
\includegraphics[scale=0.4]{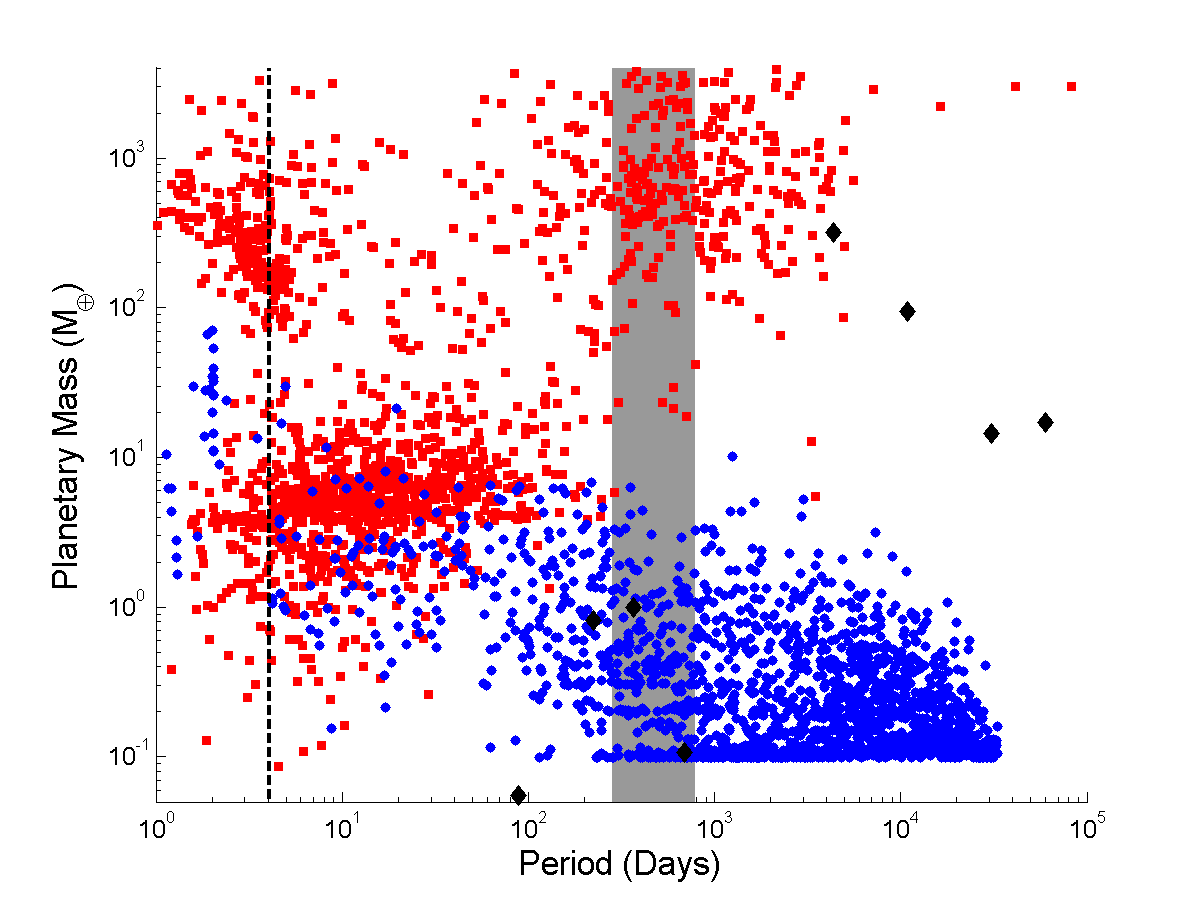}
\caption{Left panel: Mass vs period plot, comparing observed exoplanets (red squares) with our simulation results (blue circles) and the Solar System planets (black diamonds). Right panel: Same as left panel but data taken from \citet{ColemanNelson16}. The grey zones indicate the habitable zone \citep{Kasting}. The arrows at the bottom of the left panel indicate the average positions of the four radial structures.}
\label{fig:massvperiod}
\end{figure*}

\subsection{Ensemble results}
We now discuss the results of the simulations as a whole, focusing first on the masses and periods of the planets that form, and then on the eccentricity distribution.
\subsubsection{Masses and periods}
\label{sec:res_overall}
Considering the results of the simulations as a whole, 132 surviving giant planets are formed with masses ranging from $0.3\, {\rm M}_{\rm Jupiter}$ to $4\, {\rm M}_{\rm Jupiter}$, with periods from 5 days 
up to 24000 days (the smaller period being determined by our boundary conditions).
The majority of these giant planets formed at the outer edges of radial structures, whilst a handful of less massive giant planets accreted the majority of their gas envelopes after escaping from the radial structures and type I migrating towards the central star.
Fig. \ref{fig:massvperiod} shows a mass versus period diagram for all of the surviving planets from the simulations, along with all confirmed exoplanets \citep{exoplanets_org}.

The known exoplanets form three apparently distinct groups in the mass-period diagram: cold Jupiters with orbital periods $\gtrsim 100$ days; hot Jupiters with orbital periods $\lesssim 10$ days; super-Earths/Neptunes with periods between $2 \lesssim P \lesssim 100$ days. These features are affected by a number of observational biases, including the fact that ground based transit surveys are only sensitive to Jupiters with periods $\lesssim 10$ days. Nonetheless, analysis of the period distribution of planets detected only by radial velocities seems to confirm that there is a real valley in the distribution between 10--100 days \citep{Udry03,Cumming2008}. More recently, \cite{Santerne16} have presented an analysis of giant planets discovered by the \emph{Kepler} spacecraft that were followed-up using radial velocity measurements over 6 years, and they confirm that the period-valley also exists within this data set. One of the most striking features when comparing the results of the simulations with the observational data in Fig.~\ref{fig:massvperiod} is the fact that the giant planets formed in the simulations are almost all hot Jupiters (periods $< 10$ days) and cold Jupiters (periods $> 100$ days), with only a few massive bodies being located in the region that corresponds to the observed period valley. Furthermore, the simulations produce numerous planets with masses in the range $0.5 \me \lesssim m_{\rm p} \lesssim 30 \me$ and periods between $2 \lesssim P \lesssim 100$ days, that correspond to the observed super-Earths and Neptunes. Some of these lower mass planets are in systems that contain giant planets, as described in the previous section for run CJ120.1210A, and some are devoid of any giants. Comparing the left panel of Fig. \ref{fig:massvperiod} with the right panel (a reproduction of Fig. 12 in \citet{ColemanNelson16}), where similar N-body simulations, but without the inclusion of disc radial structures, were presented, we see that the agreement between the observed and simulated planet distributions is much improved in this work.

In the simulations, the origin of the two distinct populations of hot and cold Jupiters, and the period valley between them, can be explained as follows. Giant planets that form early in the disc lifetime migrate all the way into the magnetospheric cavity, and become hot Jupiters. Giant planets that are destined to \emph{not} become hot Jupiters must form near the end of the disc lifetime, when photoevaporation plays an important role in the disc evolution. Photoevaporation, combined with viscous evolution, causes the disc to disperse from the inside out. There is a high probability that a giant forming towards the end of the disc lifetime will migrate towards the star when the disc interior to the critical radius for photoevaporation has been fully or partially evacuated, preventing it from migrating close to the star, and ensuring that it remains as a cold Jupiter. Hence, the observed giant planet period distribution may arise as a combination of forming giant planets at large radius, having a stopping mechanism for migration at the inner edge of the disc (i.e. a magnetospheric cavity) and the inclusion of photoevaporation, which occurs outside a well-defined radius corresponding roughly to where a thermal wind can be launched. The influence of different models of photoevaporation on the results are discussed in more detail below, but we note that \citet{AlexanderPascucci12} have suggested that disc clearing due to photoevaporation can be responsible for a pile-up of giant planets at $1\au$, as planet migration is slowed when photoevaporation begins to dominate disc evolution. More recently \citet{ErcolanoRosotti15} showed that different models of photoevaporation influence the pile-up location, with a thermal-wind launching inner radius of 1--$2\au$ being preferred.

Low mass, compact systems that formed and migrated to the inner regions of the disc are seen in a number of simulations. The formation of these compact systems occurs similarly to those described in \citet{ColemanNelson16}, but some compact systems within this work contained giant planets with large orbital periods, as shown in sect. \ref{sec:example}.
The co-existence of long period giant planets and low mass compact systems in the simulation results seems to be in accord with the recent analysis of \emph{Kepler} data indicating the presence of long period giant planets around stars known to host compact multi-systems \citep{Uehara2016, Kipping2016}.

\subsubsection{Eccentricities of giant planets}
\begin{figure}
\includegraphics[scale=0.45]{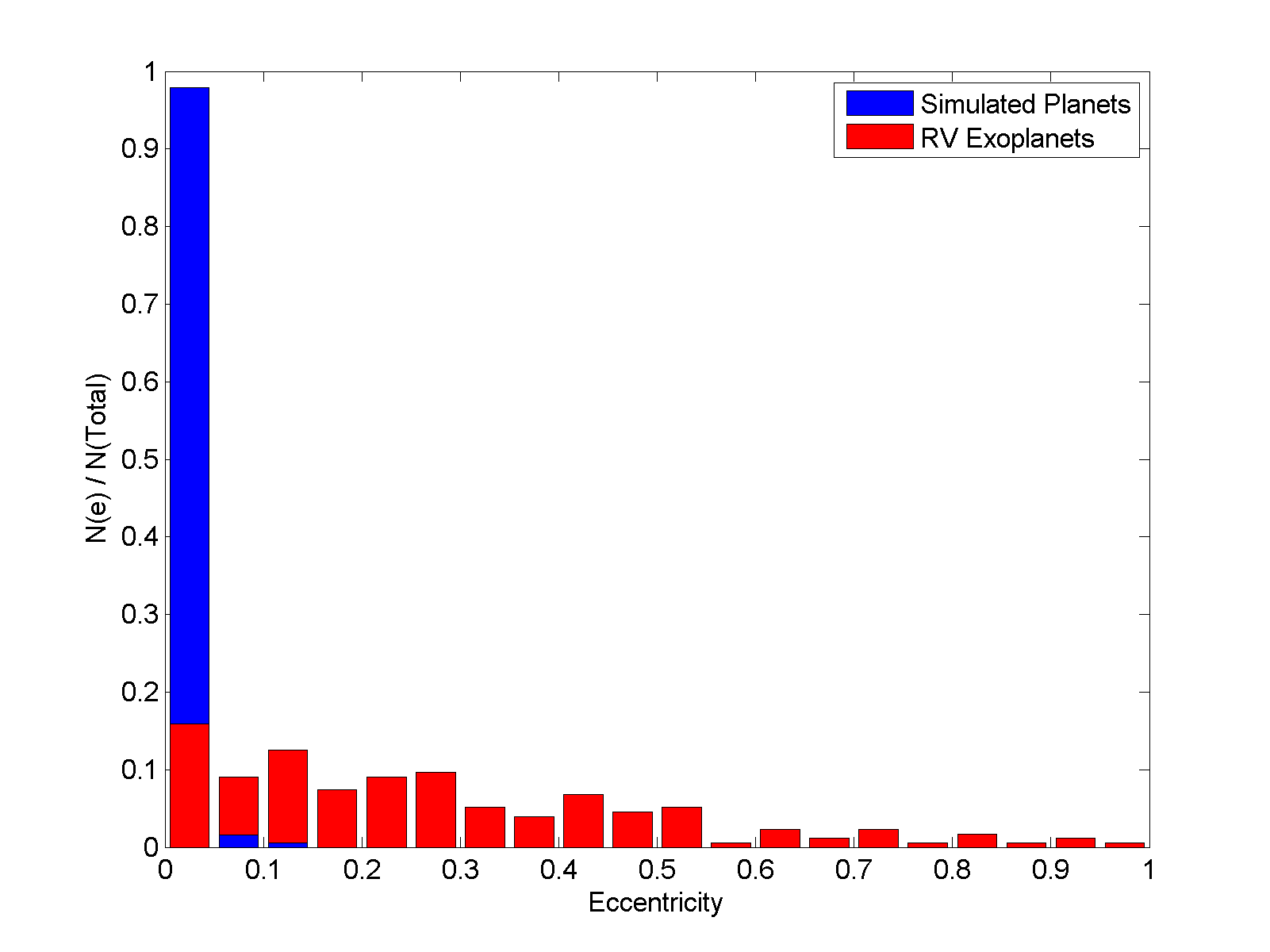}
\caption{Distribution of observed giant exoplanet eccentricities (blue) and the distribution arising from the simulations (red).}
\label{fig:ecc}
\end{figure}

The eccentricity distribution of observed giant ($m_{\rm p} \sin{i} \ge 100 \me$) exoplanets is shown in Fig.~\ref{fig:ecc} for bodies with orbital periods $>10$ days, along with the eccentricity distribution for planets in the same mass and period range that form in the simulations. It is clear that the eccentricity distribution associated with observed exoplanets is much broader than that generated in the simulations. The maximum eccentricity for a giant planet obtained in the simulations was $e_{\rm p} = 0.13$, whereas significant numbers of exoplanets are observed to have eccentricities $> 0.3$. We note that those simulated systems that resulted in modestly eccentric giants did so because the giant planets underwent strong gravitational scattering with other planets in the system, where the scattered bodies typically had masses $\simeq 20 \me$. Scattering between more massive bodies is required to obtain the larger eccentricities observed in the exoplanet data \citep[e.g.][]{RasioFord1996}.

Given that our simulations end after 10 Myr, it is possible that dynamical instabilities could occur on longer time scales in systems containing multiple giant planets, changing the statistics shown in Fig.~\ref{fig:ecc}. We have examined the distribution of mutual semimajor axis separations, expressed as a function of mutual Hill radii, to determine whether or not this is possible. We note that \citet{MarzariWeidenschilling02} examined the dynamical stability of three Jovian-mass planets on initially circular orbits, and demonstrated that the instability time scale for such a system scales with the mutual Hill radius separation, with systems separated by $\sim 6$ mutual Hill radii having instability times of $\sim 10^9$ yr. All of our systems are at least as separated as this, with approximately half of the systems having semimajor axis separations between 6 and 12 mutual Hill radii, and the other half being more separated. This suggests that some of the simulated systems may undergo dynamical instabilities on time scales longer than 10 Myr, but it seems highly unlikely that running the simulations for Gyr time scales would result in an eccentricity distribution that matches the observed one.

Assuming that the observed eccentricity distribution of giant exoplanets arises primarily because of dynamical instabilities in multiplanet systems, and using the observed distribution as a constraint on viable formation scenarios, the data suggest that giant planets must often form in compact configurations, and do so more frequently than occurs in our simulations.

Finally, we note that our simulations adopt a highly simplified prescription for the eccentricity damping experienced by gap forming planets, namely that the eccentricity is damped on a time scale of $\sim 10$ planet orbits. This is applied independently of the mass remaining in the gas disc, and so acts to bias our final systems towards having low eccentricities by reducing the likelihood of instabilities occurring while the gas disc is present. It is clear that a more sophisticated model will need to be adopted in future simulations if a more realistic assessment of the ability of the models to generate high eccentricity systems is to be undertaken.

\begin{figure}
\includegraphics[scale=0.45]{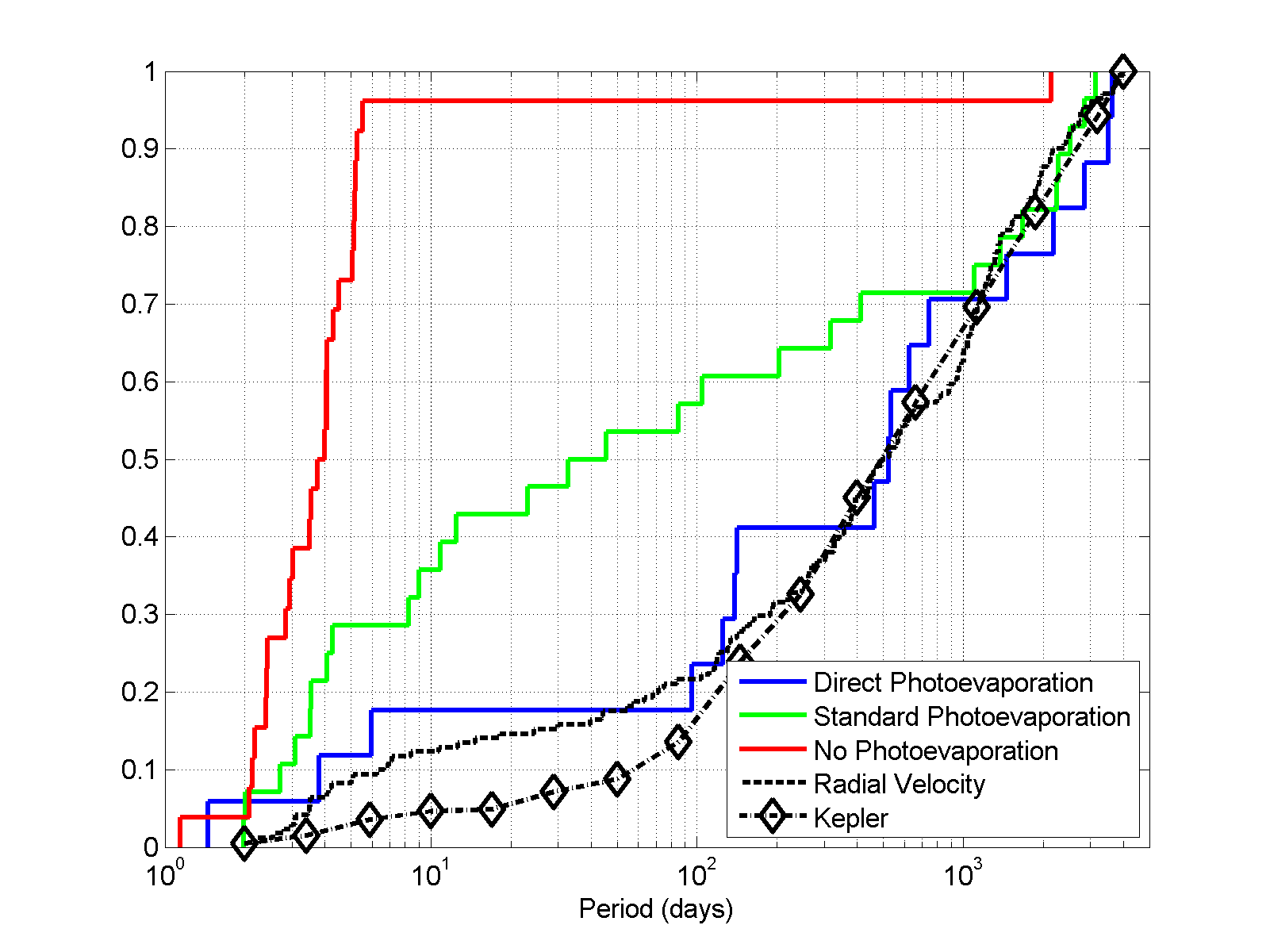}
\caption{Normalised cumulative distribution functions of giant planet periods for radial velocity (black dashed line) and \emph{Kepler} observed giant planets (black dot-dash line), and simulations with different photoevaporation regimes; direct (blue line), standard (green line), and none (red line).
We define a giant planet in both simulations and observations as a planet with mass $m_{\rm p}\ge100\me$.}
\label{fig:CDFcomp}
\end{figure}

\subsection{Different photoevaporation models}
\label{sec:res_other}

\subsubsection{Direct photoevaporation}
Simulation CJ120.1210A, presented in sect. \ref{sec:example}, was one of a group of simulations that allowed direct photoevaporation to impact the disc when the gas disc interior to the critical radius had accreted onto the central star. This can occur when a gap forming planet forms exterior to the critical photoevaporation radius, and the inner disc drains onto the star. In this scenario, the giant planet assists its own survival against migration by stimulating the onset of direct photoevaporation and reducing the disc lifetime.
Fig. \ref{fig:CDFcomp} compares the cumulative distributions of giant planet periods from simulations with different photoevaporation models (colored lines) and observations (black lines).
When comparing the observations, it is evident that for giant planets observed by \emph{Kepler}, the ratio of hot Jupiters to cold Jupiters is lower than that found by radial velocity surveys.
One possible reason for this is that the average of the metallicities of the \emph{Kepler} stars is -0.18 dex \citep{Huber14}, and this is lower than for stars in the solar neighbourhood where the average is -0.08 dex \citep{Sousa08}. 
Comparing the observations with our simulations, it is clear that the blue line, representing simulations with direct photoevaporation, compares very reasonably with the observations, albeit with a higher fraction of hot Jupiters. Given that the simulations shown here have an average metallicity of 0.3 dex, the increased ratio is perhaps unsurprising, given that the boost in solid material can allow more rapid planet formation and therefore more time for migration.
The period valley discussed above is also evident here, as is the good agreement between the simulated and observed cold Jupiter distributions.

Having observed the effect that direct photoevaporation has on the survival of giant planets with long orbital periods, we ran a further two sets of simulations with the same parameters as described in sect. \ref{sec:initial}, but with different photoevaporation models, the standard one (obtained by just switching off direct photoevaporation) and no photoevaporation, in order to examine their effects on giant planet formation.
The results of all simulations with disc mass $1\times$ MMSN, metallicity $2\times$ solar and $\alpha=2\times 10^{-3}$ are shown by the red (no photoevaporation) and green (standard) lines in Fig. \ref{fig:CDFcomp} and are discussed in the following sections.

\begin{figure*}
\includegraphics[scale=0.4]{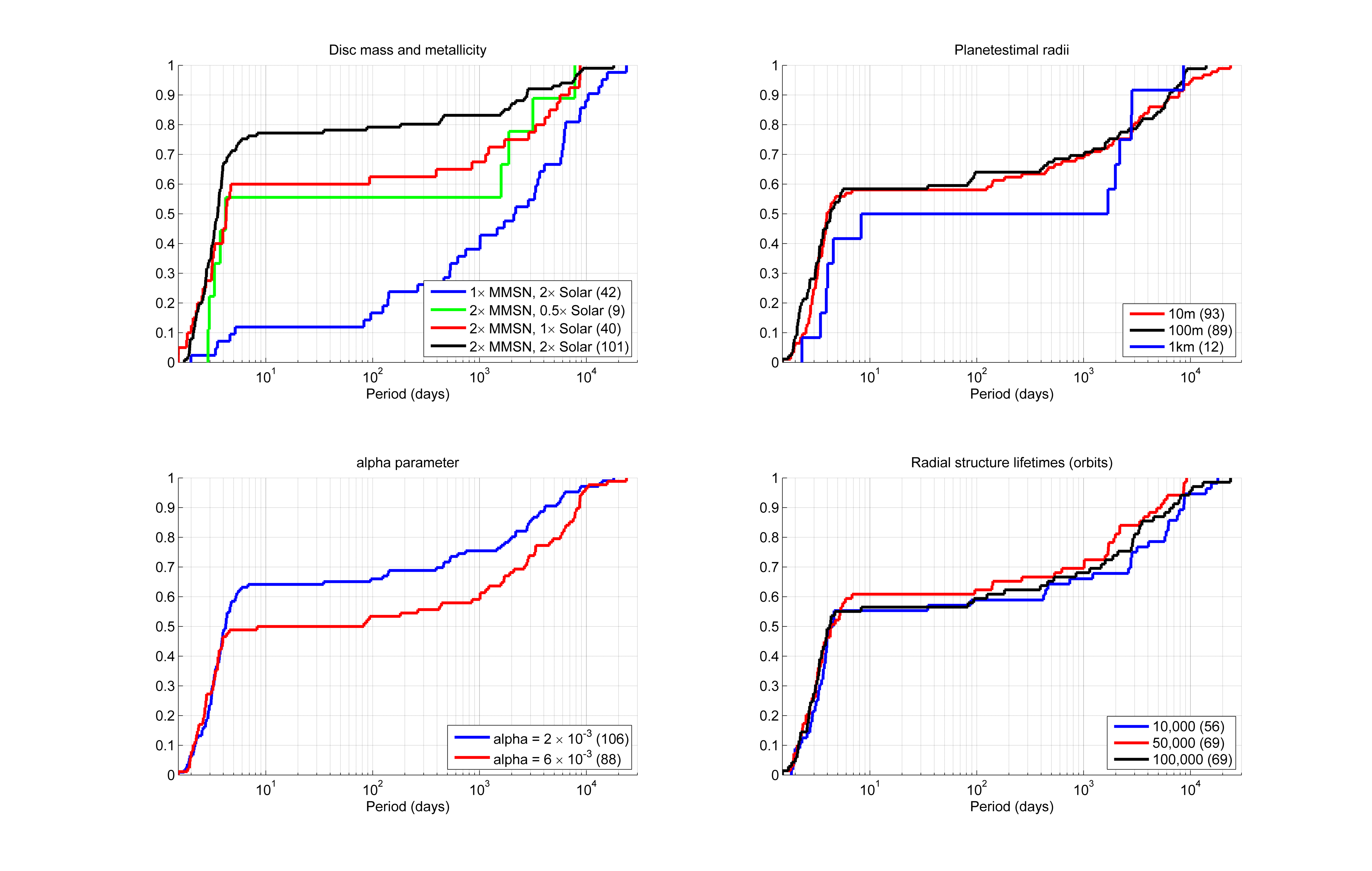}
\caption{Normalised cumulative distributions of simulated giant planets as a function of different parameters. Top-left panel: Disc mass and metallicity. Top-right panel: Planetesimal radii. Bottom-left panel: $\alpha$ parameter. Bottom-right panel: Radial structure lifetimes. Bracketed values represent the number of giant planets in those cumulative distributions. We define a giant planet in both simulations and observations as a planet with mass $m_{\rm p}\ge100\me$.}
\label{fig:allCDFs}
\end{figure*}

\subsubsection{Standard photoevaporation}
\label{sec:res_diffuse}
Given that only a modest number of our simulations containing large $\gtrsim 1$~km planetesimals formed giant planets, we only ran simulations with 10~m boulders or 100~m planetesimals to examine the influence of switching off direct photoevaporation and retaining the standard photoevaporation model. We note that when comparing the results of simulations that employed standard and direct photoevaporation models, evolution of the disc and planets are identical until the time that direct photoevaporation is activated. This means that the formation pathways of giant planets is similar, and significant differences only arise for those cases where giant planet formation and migration occurs near to the end of the disc lifetime, when photoevaporation is strongly influencing the disc evolution. Direct photoevaporation causes the disc to be removed more rapidly, and so is more effective at stranding migrating planets at larger orbital radii. Planets that form and migrate in discs with standard photoevaporation are therefore more likely to form hot Jupiters, as indicated in Fig. \ref{fig:CDFcomp}.

\subsubsection{No photoevaporation}
\label{sec:res_none}
In this set of simulations, we neglect photoevaporation entirely, such that the only processes that can deplete the gas disc are viscous evolution and accretion onto planets, significantly increasing disc lifetimes and the time periods over which migration can occur. We consider only models containing 10~m boulders and 100~m planetesimals. 

The early formation and evolution of giant planets is similar to that seen in simulations with photoevaporation. Once a giant planet forms, however, the lack of an effective disc removal mechanism means that it will almost always migrate all the way to becoming a hot Jupiter,
as shown by the red line in Fig. \ref{fig:CDFcomp}, where 95\% of the giant planets formed are hot Jupiters. The giant planets that remain as cold Jupiters only did so because they formed late in the disc lifetime, where they survived migration by accreting the majority of the remaining gas disc.
This ratio of hot Jupiters to cold Jupiters is not consistent with observations, and shows that recreating the observed distributions of giant planets is extremely difficult without a mechanism for disc dispersal.

\subsection{Evolution as a function of model parameters}
We now discuss the effects that varying the model parameters have on the formation and evolution of giant planets in the simulations. Since these effects are consistent across all photoevaporation models employed, we will only discuss the simulations that include direct photoevaporation.
Fig. \ref{fig:allCDFs} shows the cumulative distributions for simulated planets as a function of the different parameters considered.

\subsubsection{Disc mass and metallicity}
The simulation results show a strong dependence on the initial mass and metallicity of the disc.
Simulations with small disc masses and sub-solar metallicities (e.g. $1\times$ MMSN and $0.5 \times$ solar metallicity) are unable to form any giant planets, due to the quantity of solid material in the disc being insufficient to form a massive planet core capable of accreting a massive gas envelope during the disc life time. Increasing the inventory of solids by increasing the total disc mass, or by increasing the metallicity, leads to the formation of giants. We see from Fig. \ref{fig:allCDFs}
that the $1\times$ MMSN, $2\times$ solar metallicity runs  form moderate numbers of hot Jupiters,
with 90\% of the giant planets having periods $>100$ days. This is for the following reasons: the planet cores form quite late in the disc lifetime; the disc lifetime is shorter than for heavier discs; the low disc mass leads to slower type I migration. Increasing the disc mass and metallicity can be seen to dramatically increase the numbers of hot Jupiters, as planet cores form earlier, type I migration is faster and the disc lifetime is longer. Models with disc mass $2\times$ MMSN and metallicity $2\times$ solar form numerous giant planets, and 80\% of these are hot Jupiters. 

\subsubsection{Planetesimal radii}
The cumulative distributions for the giant planet orbital periods formed in simulations with different planetesimal radii are shown in the top-right panel of Fig.~\ref{fig:allCDFs}.
No giant planets formed in simulations where the planetesimal size was 10~km, in agreement with the very anaemic growth found in \cite{ColemanNelson16} for models with 10~km planetesimals. Large planetesimals do not migrate very far through the disc during its lifetime, and the relatively weak damping means that their accretion rate onto planetary embryos remains small because of their large velocity dispersion. Accretion rates are slightly higher for 1~km planetesimals, leading to 12 giant planets forming in these runs. Overall, only $\sim 5\%$ of all giant planets formed do so in simulations with 1 or 10~km sized planetesimals (half of all runs). When the planetesimal radius is decreased to 100~m, or we consider 10~m boulders, then giant planets form easily. \cite{ColemanNelson16} found that planetary growth is efficient in the presence of small bodies that experience strong gas drag, since they can migrate over large distances (helping growing embryos to exceed their local isolation masses), and maintain a relative modest velocity dispersion due to strong eccentricity and inclination damping. Our inclusion of radial structures allows small planetesimals and boulders to concentrate, and growing embryos to avoid rapid inward migration. Hence, the simulations form surviving giant planets with a broad range of orbital periods. Similar numbers of giant planets formed in simulations with 10~m and 100~m small bodies, while their orbital period distribution (i.e. number of hot Jupiters versus cold Jupiters) was also similar, as is shown by the cumulative distributions in the top-right panel of Fig. \ref{fig:allCDFs}.

\subsubsection{$\alpha$ viscosity}
The bottom-left panel of Fig. \ref{fig:allCDFs} shows that a lower viscosity (i.e. mass accretion rate through the disc for a given disc mass) gives rise to a larger ratio of hot to cold Jupiters. This is an effect of the shorter disc lifetimes associated with more viscous discs, by approximately 2 Myr in our simulations. A closely related effect is that the numbers of giant planets that form in higher viscosity discs is lower than in lower viscosity discs: 88 formed in the $\alpha=6\times 10^{-3}$ runs versus 106 in the $\alpha=2 \times 10^{-3}$ simulations.

\subsubsection{Radial structure lifetime}
The cumulative distributions of orbital period for runs with different assumed lifetimes for the disc radial structures are shown in the bottom-right panel of Fig. \ref{fig:allCDFs}.
It is clear that varying these lifetimes between $10^4$ and $10^5$ local orbit periods has very little influence on the results. We expect that shorter lifetimes than those considered in the runs would reduce the numbers of giant planets that form, since more growing cores could escape from the outer disc regions and migrate rapidly into the inner magnetospheric cavity before becoming giants. It is also likely that the mass distribution of the giants would be skewed towards lower masses, and the ratio of hot to cold Jupiters would increase. By decreasing the lifetimes of the radial structures to very short values we would eventually converge towards the results presented in \citet{ColemanNelson16}, where all surviving giant planets were hot Jupiters and had sub-Jovian masses.

\begin{figure*}
\includegraphics[scale=0.73]{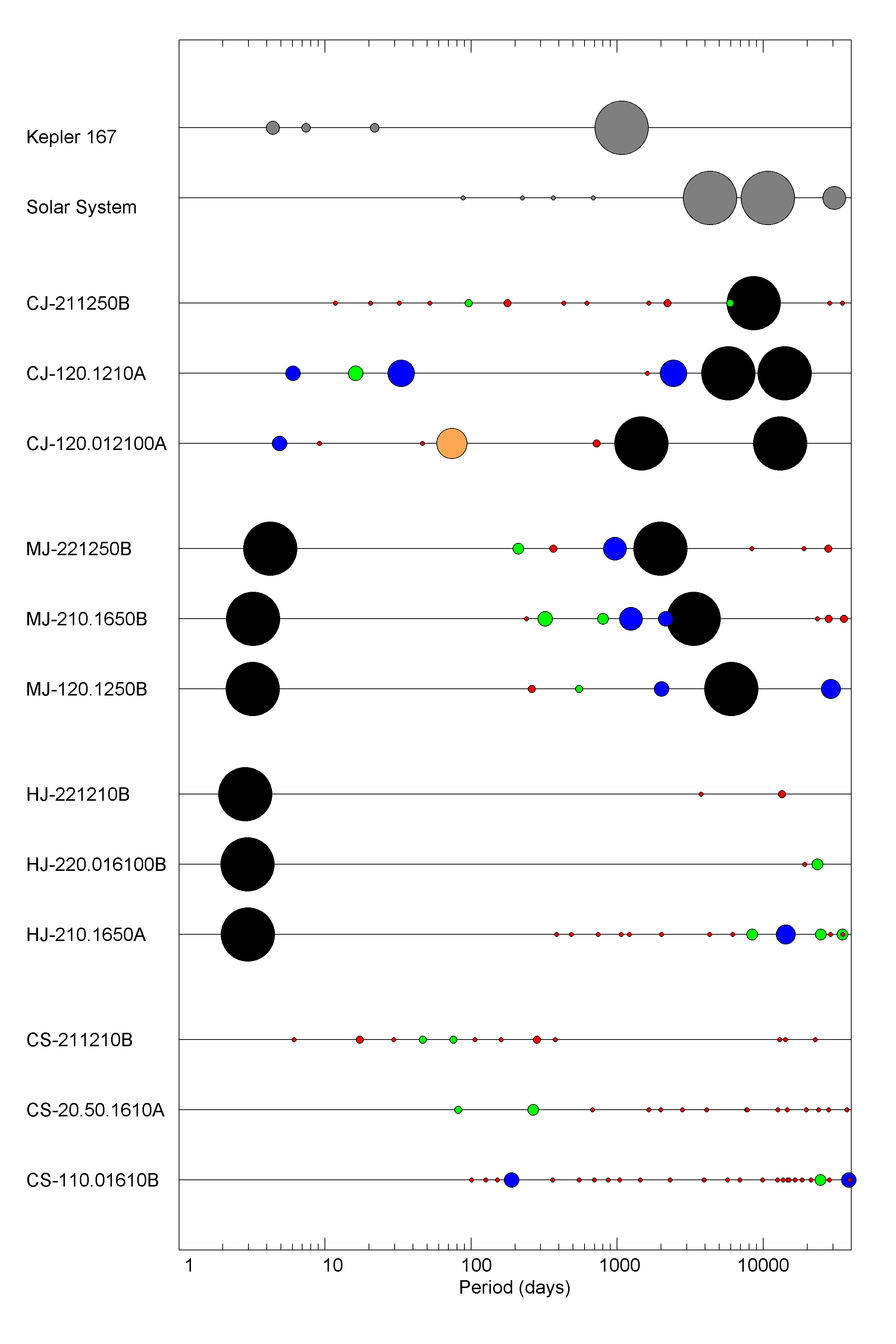}
\caption{Plot comparing different architectures arising from the simulations, with the Solar System and Kepler-167 included for comparison.
Orbital period is indicated on the $x$-axis and planet
masses are indicated by the symbol size (radius scales 
with the square-root of the planet mass).
The symbol colours indicate the classification of each planet:
red = Earths ($m_{\rm p}<3\me$); green = super-Earths ($3\me\le m_{\rm p}<10\me$); blue = Neptunes ($10\me\le m_{\rm p}<35\me$); orange = super-Neptunes ($35\me\le m_{\rm p}<100\me$); black = Jupiters and super-Jupiters ($m_{\rm p}>100\me$). See Table~\ref{tab:classifications} for definitions of planet types.}
\label{fig:compsystems}
\end{figure*}

\subsection{Planetary system architectures}
We find a diversity in the planetary system architectures arising from the simulations.
An ensemble of simulated planetary systems displaying different architectures are shown in Fig. \ref{fig:compsystems}, where the different architectures are represented by different simulation label prefixes.
Below we describe the different architectures, and the general physical conditions and modes of evolution associated with each of them:\\
(i) \emph{Low-mass planetary systems} -- These form in simulations where protoplanet growth rates are insufficient to form giant planets.
In some cases these are compact planetary systems, with similar formation histories to those discussed in \citet{ColemanNelson16}.
The systems with the prefix `CS' (compact system) in Fig. \ref{fig:compsystems} show the final configurations from these runs, where the lack of massive planets is evident along with their compactness.
Generally, these systems arose in metal-poor low-mass discs with small planetesimals/boulders, or in more massive discs with large planetesimals (i.e. $R_{\rm pl}\ge1$~km).\\
(ii) \emph{Lonely hot Jupiters} -- Systems containing only hot Jupiters formed in massive metal-rich discs. Typically multiple giant planets form in the outer regions of the disc and migrate to become hot Jupiters, where only the last hot Jupiter survives.
Often accompanying these hot Jupiters are  low-mass planets on long period orbits ($P_{\rm p}\ge 100$d), as shown by systems with prefixes `HJ' (hot Jupiters) in Fig. \ref{fig:compsystems}.
From an observational perspective, the low mass and long orbital periods of these companions would make the hot Jupiters appear singular.\\
(iii) \emph{Hot Jupiters with cold Jupiter companions} -- Similar to the \emph{lonely hot Jupiters}, planetary systems that contain both hot and cold Jupiters tend to arise from solids-rich discs.
Hot Jupiters form early in the disc lifetime and migrate close to the central star, whilst late forming giant planets have insufficient time to migrate into the inner system, retaining long orbital periods as cold Jupiters. Typically, lower mass planets are found to occupy the space between the hot and cold Jupiters. Examples of these systems are shown in Fig. \ref{fig:compsystems} by the prefix `MJ' (multiple Jupiters), showing the diversity in planetary compositions in these systems.\\
(iv) \emph{Cold Jupiters with low-mass companions} -- When there is sufficient solid material in the disc, we find that giant planets can form simultaneously with inner systems of low mass planets.
The late formation of giant planets enables them to remain as cold Jupiters at the end of the disc lifetime, whilst interior low-mass planets slowly accrete and migrate into the inner disc regions, becoming an inner system of low-mass planets, occasionally in a compact configuration.
This architecture is similar to that found in the Solar System, and we note that recent analysis of \emph{Kepler} light-curves indicates the existence of long period giant planets orbiting stars with known compact low mass systems, similar to our simulated cold Jupiters with short period low mass companions \citep{Uehara2016,Kipping2016}.
This planetary system architecture is shown by systems with the prefix `CJ' (cold Jupiters) in Fig. \ref{fig:compsystems}. \\

\section{Discussion and conclusions}
\label{sec:discussion}
We have presented the results of N-body simulations coupled with prescriptions for planetary migration, accretion of gaseous envelopes, self-consistent evolution of a viscous disc with an inner magnetospheric cavity and disc removal by a photoevaporative wind on multi-Myr time scales. A new addition, not considered in our previous simulations \citep{ColemanNelson14, ColemanNelson16}, is radial structuring of the disc due to variations in the viscous stresses, leading to the formation of persistent planet traps at large orbital radii from the star. The main results from our study can be summarised as follows: \\
(i) Radial structuring of the disc allows gas giant planets to form. Protoplanets and planetesimals become trapped at the outer edges of the radial structures, due to strong corotation torques and positive pressure gradients, respectively. Giant planet cores capable of accreting gaseous envelopes are able to form due to efficient accretion of planetesimals/boulders by planetary embryos.
Out of 288 simulations, 132 surviving giant planets were formed by having their cores trapped by radial structures. The final periods depend on the time and location of formation, as discussed in \citet{ColemanNelson14}, where early forming giant planets became hot Jupiters, and late forming giant planets remain as cold Jupiters.\\
(ii) When analysing the effects of changing specific parameters, we identify the following trends:\\
-- In solid-poor simulations (low disc mass and metallicity) no giant planets are formed, as there is insufficient solid material to form giant planet cores. This is in agreement with the observations of \citet{FischerValenti05} and \citet{Santos04}, where giant planets are preferentially found around metal-rich stars. \\
-- When the planetesimal radii are large ($\ge 1$km), giant planets are unable to form except in the most solids-rich environments. Giant planet formation is strongly favoured in models where the primary feedstock of planetary growth is in the form of small 100~m sized planetesimals or 10~m sized  boulders.
$95\%$ of the giant planets that formed did so in simulations with small boulders/planetesimals. None were formed in models with 10~km planetesimals.\\
-- We find that discs with higher viscosity form fewer giant planets than low viscosity discs, and the ratio of hot to cold Jupiters in higher viscosity discs is smaller than in lower viscosity discs. These effects are entirely due to the shorter disc lifetimes associated with higher viscosity.\\
(iii) Multiple giant planets are able to form when there is sufficient solid material. This occurred in numerous simulations with high disc masses and metallicities, resulting in systems with multiple cold Jupiters, or a hot Jupiter with cold Jupiter companions. The survival rate of warm Jupiters (those with periods between 10--100 d) also depends on the presence of outer giant companions. Outer giant planet companions can stem the flow of gas into the inner system, reducing the migration rate of planets in the inner system and allowing them to survive at longer periods than if there were no exterior giant planets.\\
(iv) Our simulations reproduce the  giant planet period valley between 10 and 100 days that is seen in the observed period distribution of giant planets. Our analysis shows that this arises because of the inclusion of disc removal by photoevaporation in our simulations. The launching of a photoevaporative wind causes the disc to empty from the inside out at the end of its lifetime, causing the migration of planets to stall at periods $> 100$ days, an effect that has been discussed previously by \citet{AlexanderPascucci12} and \citet{ErcolanoRosotti15}.\\
(v) Our simulations do not reproduce the broad eccentricity distribution of the observed giant exoplanets, and this is apparently because our multiple giant planet systems are too well separated to undergo dynamical instabilities that lead to the formation of eccentric orbits. We note, however, that our use of a simple model for damping the eccentricity of gap forming planets in the presence of the gas disc may also bias the simulations towards producing low eccentricity systems. A definitive conclusion about the ability of our models to form a population of eccentric giants can only be made once an improved prescription for this has been implemented. A further point that is worth making is that systems of multiple giant planets form in the simulations when the system metallicity is high (as described above). Assuming that the primary mechanism leading to the observed giant exoplanets attaining their eccentric orbits was dynamical instability in multiplanet systems (possibly on time scales much longer than the formation time scales that we have considered), we note that this (not unexpected) correlation between metallicity and the multiplicity of giant planets that form in the simulations may also explain the positive correlation that exists between eccentricity and stellar metallicity for giant exoplanets discovered by radial velocity surveys. \footnote{This correlation may be seen by plotting eccentricity versus stellar metallicity using the data on radial velocity planets available at exoplanets.org}
We note that this correlation has also been pointed out by \cite{Dawson13}.\\
(vi) Numerous compact systems of super-Earths and Neptunes were formed in the simulations, with formation histories similar to those discussed in \citet{ColemanNelson16}. If there was sufficient solid material, long period giant planets also formed in the same simulations as the compact systems of super-Earths/Neptunes.

The simulations we have presented here show that giant planets can form in discs containing radial structures that act as planet traps, while the combination of magnetospheric cavities and photoevaporative winds creates two populations of giant planets: hot Jupiters and cold Jupiters.
It is likely that in more realistic discs, the location, size and evolution of radial structures will be quite different from what we have examined in this work. Running a full parameter study on the effects of radial structures in protoplanetary discs, however, goes beyond the scope of this study, which is intended to be a proof of concept rather than an exhaustive survey of parameter space.

Whilst this and recent work \citep{ColemanNelson16} can more or less explain some of the diversity observed in exoplanetary system architectures, it is by no means complete.
Further improvements to the model are required to enhance the accuracy and realism currently provided by simple assumptions. In future work we will aim to include the following improvements:\\
(i) A full collision and fragmentation model, so that the outcomes of planet-planet and planet-planetesimal collisions can be accurately modelled, instead of the current assumption of perfect mergers. In addition to accounting for effects arising from impacts between solid bodies, it will also be important to incorporate a model for the post-collision evolution of gaseous envelopes \citep{Liu2015a, Liu2015b}. \\
(ii) Incorporating a more realistic migration model that takes into account 3D effects \citep{Fung2015}, the influence of planet luminosity \citep{Benitez-LlambayMasset2015} and dynamical torques arising from the planet's migration \citep{Paardekooper2014, Pierens15}.\\
(iii) Calculation of gas envelope accretion using self-consistent calculations that include the effects of changing local disc conditions and planetesimal accretion rates, rather than using fits to the \citet{Movs} models that strictly apply only to planets at fixed locations in a non-evolving disc with specifically prescribed planetesimal accretion rates.\\
(iv) Incorporating fits to MHD simulations so that disc radial structures arising from zonal flows and transitions between magnetically active and dead zones can be included in a more realistic fashion. \\
(v) Incorporating an improved model for the eccentricity evolution of gap forming planets. At present the model simply assumes eccentricity damping on a time scale of $\sim 10$ orbital periods, independent of the mass contained in the disc (although the damping is removed after disc removal). This biases our systems towards low eccentricity, and prevents us from properly assessing whether or not the simulations that we have presented are able to reproduce the eccentricity distribution of the observed giant exoplanets.

Finally, it is important to emphasise that this study is not an exercise in population synthesis, as we have not chosen initial conditions (e.g. disc masses, disc lifetimes) from observationally motivated distribution functions. Instead we have simply sought to examine what conditions are required to form giant planets whose gross orbital characteristics are similar to the observed distribution. Once our model becomes more sophisticated we will examine whether or not it is capable of reproducing the observed distribution of exoplanets with appropriate choices of initial conditions.

\bibliographystyle{mnras}
\bibliography{references}{}

\end{document}